\documentclass[12pt,onecolumn]{IEEEtran}
	\usepackage{graphicx,amssymb,amsmath}
	\usepackage{multicol}
	\usepackage[noadjust]{cite}
	\usepackage{setspace}
	\usepackage{subfigure}
	\usepackage{graphicx}
	\usepackage{float}
	\usepackage {url}
	\usepackage{stfloats}
	\usepackage{amsthm,pifont}
	\usepackage{flushend}
	\usepackage{cases,subeqnarray}
	\usepackage{bm,multirow,bigstrut}
	\usepackage{amsmath, amsthm, amssymb}
	\usepackage{textcomp}
	\usepackage{latexsym,bm}
	\usepackage{booktabs}
	\usepackage{xcolor}
	\usepackage{mathtools}
	\usepackage{dsfont}
	\usepackage{extarrows}
	\usepackage{epsfig}
	\usepackage{epsfig}
	\usepackage{epstopdf}
	\usepackage[noend]{algpseudocode}
	\usepackage{algorithmicx,algorithm}
	\theoremstyle{plain}

	\theoremstyle{plain}
	\newtheorem{rem}{Remark}
	\newtheorem{them}{Theorem}
	\newtheorem{corr}{Corollary}
	\newtheorem{prop}{Proposition}
	\usepackage{amsmath}
	
	\newtheorem{lemma}{Lemma}
	\IEEEoverridecommandlockouts
	
\begin{document}
	\title{Millimeter Wave Communications With Reconfigurable Intelligent Surfaces: Performance Analysis and Optimization}
	\author{Hongyang Du, Jiayi~Zhang,~\IEEEmembership{Senior~Member,~IEEE}, Julian Cheng,~\IEEEmembership{Senior~Member,~IEEE}, \\ and Bo Ai,~\IEEEmembership{Senior~Member,~IEEE}
	\thanks{H. Du and J.~Zhang are with the School of Electronic and Information Engineering, Beijing Jiaotong University, Beijing 100044, China. (e-mail: \{17211140; jiayizhang\}@bjtu.edu.cn)}
	\thanks{J. Cheng is with the School of Engineering, The University of British Columbia, Kelowna, BC V1V 1V7, Canada. (e-mail: julian.cheng@ubc.ca).}
	\thanks{B. Ai is with the State Key Laboratory of Rail Traffic Control and Safety, Beijing Jiaotong University, Beijing 100044, China. (e-mail: boai@bjtu.edu.cn)}
	}
	\maketitle
	\vspace{-1cm}
	\begin{abstract}
	Reconfigurable Intelligent Surface (RIS) can create favorable multipath to establish strong links that are useful in millimeter wave (mmWave) communications. While previous works used Rayleigh or Rician fading, we use the fluctuating two-ray (FTR) distribution to model the small-scale fading in mmWave frequency. First, we obtain the statistical characterizations of the product of independent FTR random variables (RVs) and the sum of product of FTR RVs. For the RIS-aided and amplify-and-forward (AF) relay systems, we derive exact end-to-end signal-to-noise ratio (SNR) expressions. To maximize the end-to-end SNR, we propose a novel and simple way to obtain the optimal phase shifts at the RIS elements. The optimal power allocation scheme for the AF relay system is also proposed. Furthermore, we evaluate important performance metrics including the outage probability and the average bit-error probability. To validate the accuracy of our analytical results, Monte-Carlo simulations are subsequently conducted to provide interesting insights. It is found that the RIS-aided system can achieve the same performance as the AF relay system with low transmit power. More interestingly, as the channel conditions improve, the RIS-aided system can outperform the AF relay system using a smaller number of reflecting elements.
	\end{abstract}
	\IEEEpeerreviewmaketitle
	\section{Introduction}
	As a promising technique for supporting skyrocket data rate in the fifth-generation (5G) cellular networks, millimeter wave (mmWave) communications have received an increasing attention due to the large available bandwidth at mmWave frequencies \cite{zhang2020Prospective}. In recent years, research on channel modeling for mmWave wireless communications has been intense both in industry and academia \cite{rappaport2015wideband,romero2017fluctuating}. {\color{black}Based on recent small-scale fading measurements of the $28$ {\rm GHz} outdoor millimeter-wave channels \cite{romero2017fluctuating}, the fluctuating two-ray (FTR) fading model has been proposed as a versatile model that can provide a much better fit than the Rician fading model.}

	However, one of the fundamental challenges of mmWave communication is the susceptibility to blockage effects, which can occur due to buildings, trees, cars, and even human body. To address this problem, a typical solution is to add new supplementary links. For example, the amplify-and-forward (AF) relay can be introduced to amplify the weak signal and re-transmit it toward the destination \cite{rankov2007spectral}. Alternatively, reconfigurable intelligent surfaces (RISs), comprised of many reflecting elements, have recently drawn significant attention due to their superior capability in manipulating electromagnetic waves \cite{cui2014coding}. Taking advantage of cheap and nearly passive RIS attached in facades of buildings, signals from the base station (BS) can be re-transmitted along desired directions by tuning their phases shifts, thereby leveraging the line-of-sight (LoS) components between the RIS and users to maintain good communication quality.
	
	Obviously, RIS and relay operate in different mechanisms to provide supplementary links. With the help of RIS, the propagation environment can be improved because of extremely low-power consumption without introducing additional noise, but the incident signal at the reflector array is reflected without being amplified. Thus, it is of interest to compare RIS and relay in term of efficient or cost. In \cite{huang2019reconfigurable}, a comparison between RIS and an ideal full-duplex relay was made and it was found that large energy efficiency gains by using an RIS, but the setup is not representative for a typical relay. Besides, authors in \cite{bjornson2019intelligent} made a fair comparison between RIS-aided transmission and conventional decode-and-forward (DF) relaying, with the purpose of determining how large an RIS needs to be to outperform conventional relaying, and it is found that a large number of reflecting elements are needed to outperform the DF relaying in terms of minimizing the total transmit power and maximizing the energy efficiency. However, previous works did not build on a versatile statistical channel model that well characterizes wireless propagation in mmWave communications to derive performance metrics.
	
	In this paper, we aim to answer the significant question \textit{``How can a RIS outperform AF relaying over realistic mmWave channels?"}. Using the FTR fading channel model, we derive novel exact expressions to analyze the system performance for both systems. The main contributions of this paper are summarized as follows:
	\begin{itemize}
	\item We derive the exact probability density function (PDF), cumulative distribution function (CDF), generalized moment generating function (MGF) of a product of independent but not identically distributed (i.n.i.d.) FTR random variables (RVs) and the sum of product of FTR RVs. These statistical characteristics are useful in many communication scenarios, such as multi-hop communication systems \cite{hasna2003outage} and keyhole channels of multiple-input multiple-output (MIMO) systems \cite{shin2004performance}.
	
	\item We propose an optimal phase shift design method using a binary search tree algorithm to obtain RIS's reflector array. This method does not require to estimate channel model. Furthermore, the convergence of the proposed phase optimization method is investigated. Our algorithm can guarantee to converge to the optimal solution even when the size of the RIS is large. By increasing number of iterations, satisfactory accuracy can be achieved upon convergence for any initialization. To provide a fair comparison between RIS-aided transmission and AF relaying, we further propose an optimal power allocation scheme for AF relay systems.
	
	\item We derive a novel generic single integral expression for the CDF of the end-to-end signal-to-noise ratio (SNR) of AF systems by considering not identically distributed hops and hardware impairments. With the help of the obtained statistics of FTR RVs, we derive the exact PDF and CDF of the end-to-end SNR for RIS-aid system and AF relay system.
	
	\item We provide a fair comparison between RIS-aided  system and AF relay aided communication system. The exact outage probability (OP) and average bit-error probability (ABEP) expressions are derived for both scenarios to obtain important engineering insights. It is interesting to discover that the RIS-aided system can achieve the same OP and ABEP as the AF relay system with less reflecting elements if the transmit power is low. More importantly, as the channel conditions improve, the RIS-aided system achieves more ABEP reduction than the AF relay aided system having the same transmit power.
	\end{itemize}
	
	The remainder of the paper is organized as follows. In Section \ref{sysmodel}, we briefly introduce the RIS-aided and AF relay systems, and derive exact statistics for the end-to-end SNR of both systems. Section \ref{optimals} presents the optimal phase shift of the RIS's reflector array and the optimal power allocation scheme for the AF relay system. In Section \ref{perfos}, Performance metrics of two systems, such as OP and ABEP, are derived. Moreover, Monte-Carlo simulations are presented. Finally, Section \ref{cons} concludes this paper.
	
	{\color{black}{\it Mathematical notations:} A list of mathematical symbols and functions most frequently used in this paper is available in Table \uppercase\expandafter{\romannumeral1}.}
	\begin{table}[htbp]
	{\color{black}\caption{Mathematical Symbols and Functions}}
	\centering
	{\color{black}\begin{tabular}{l|p{13cm}}
	\toprule
	$i$ &$i = \sqrt{-1}$.\\
	$\gamma$ &The squared FTR RV.\\
	$R$ &The FTR RV.\\
	$X$ &The product of FTR RVs.\\
	$Y$ &The sum of product of FTR RVs.\\
	$K$ & The average power ratio of the dominant wave to the scattering multipath in FTR fading model.\\
	$m$ &The fading severity parameter in FTR fading model.\\
	$\Delta $&A parameter varying from $0$ to $1$ representing the similarity of two dominant waves in FTR fading model.\\
	$\sigma$&The standard deviation of the diffuse received signal component in FTR fading model.\\
	$ \upsilon $&The received average SNR in FTR fading model, and $ \upsilon =2 \sigma^{2}(1+K)$\\
	$\gamma \left( \cdot, \cdot \right)$& The incomplete gamma function \cite[eq. (8.350.1)]{gradshteyn2007}.\\
	$P_ \cdot ^ \cdot \left(  \cdot  \right)$ & The Legendre function of the first kind \cite[eq. (8.702)]{gradshteyn2007}.\\
	$P$ &The transmit power of BS when RIS is used.\\
	$P_1$ &The transmit power of BS when AF relaying is used.\\
	$P_2$ &The transmit power of AF relay.\\
	$s$ &Transmit signal satisfying $ {\mathbb E}\left[ {{{\left| s \right|}^2}} \right] = 1 $\\
	$n $ & The additive white Gaussian noise (AWGN) at the user, and $n \sim {\cal C}{\cal N}\left( {0,{o ^2}} \right)$\\
	\bottomrule
	\end{tabular}}
	\end{table}
	
	\section{System Model And Preliminaries}\label{sysmodel}
	\subsection{System Description}
	\begin{figure}[th]
	\begin{minipage}[t]{0.45\linewidth}
	\centering
	\includegraphics[width=1\textwidth]{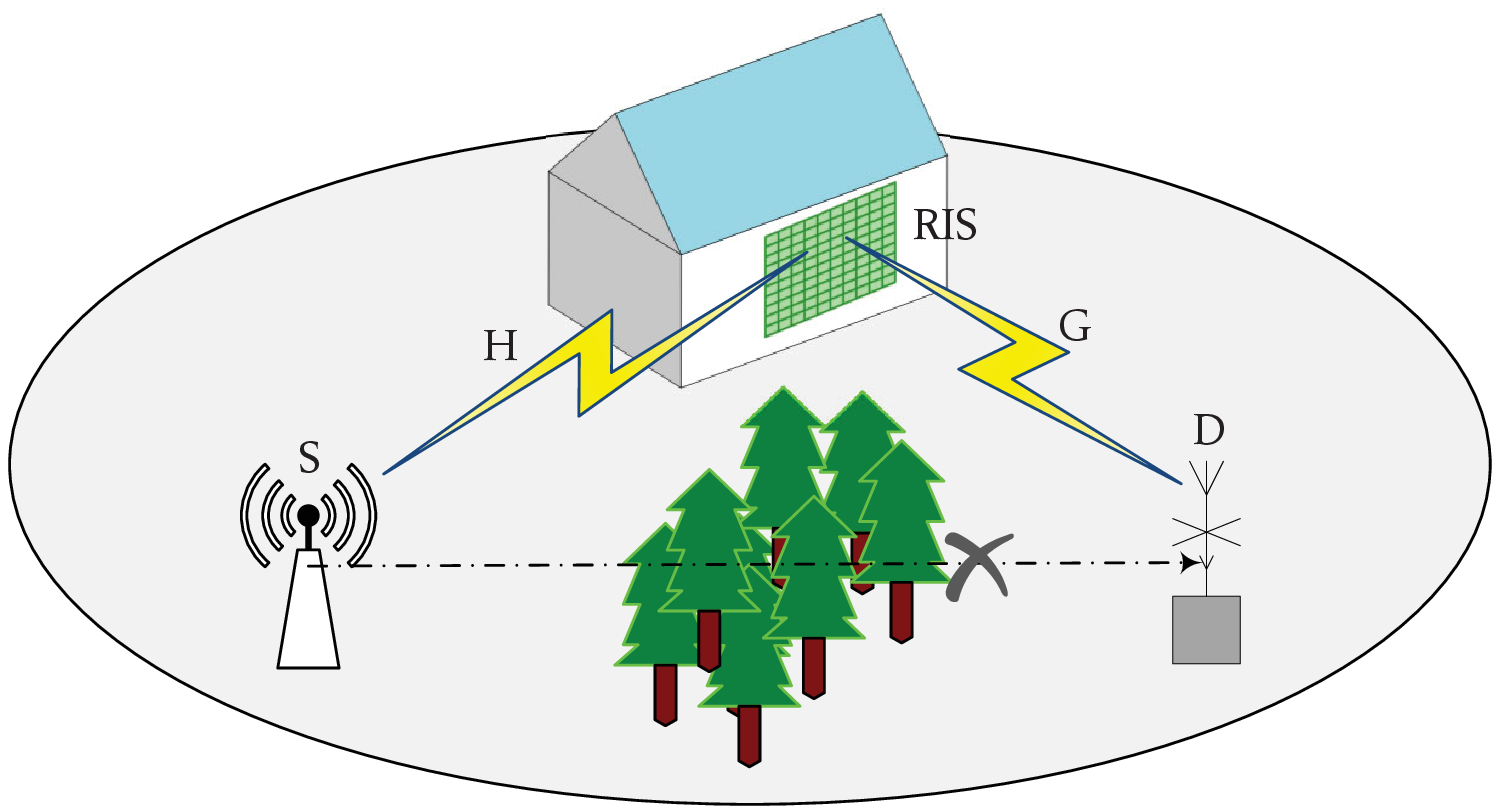}	
	\caption{Intelligent reflecting surface supported transmission.}
	\label{model1}
	\end{minipage}
	\hfill
	\begin{minipage}[t]{0.45\linewidth}
	\centering
	\includegraphics[width=1\textwidth]{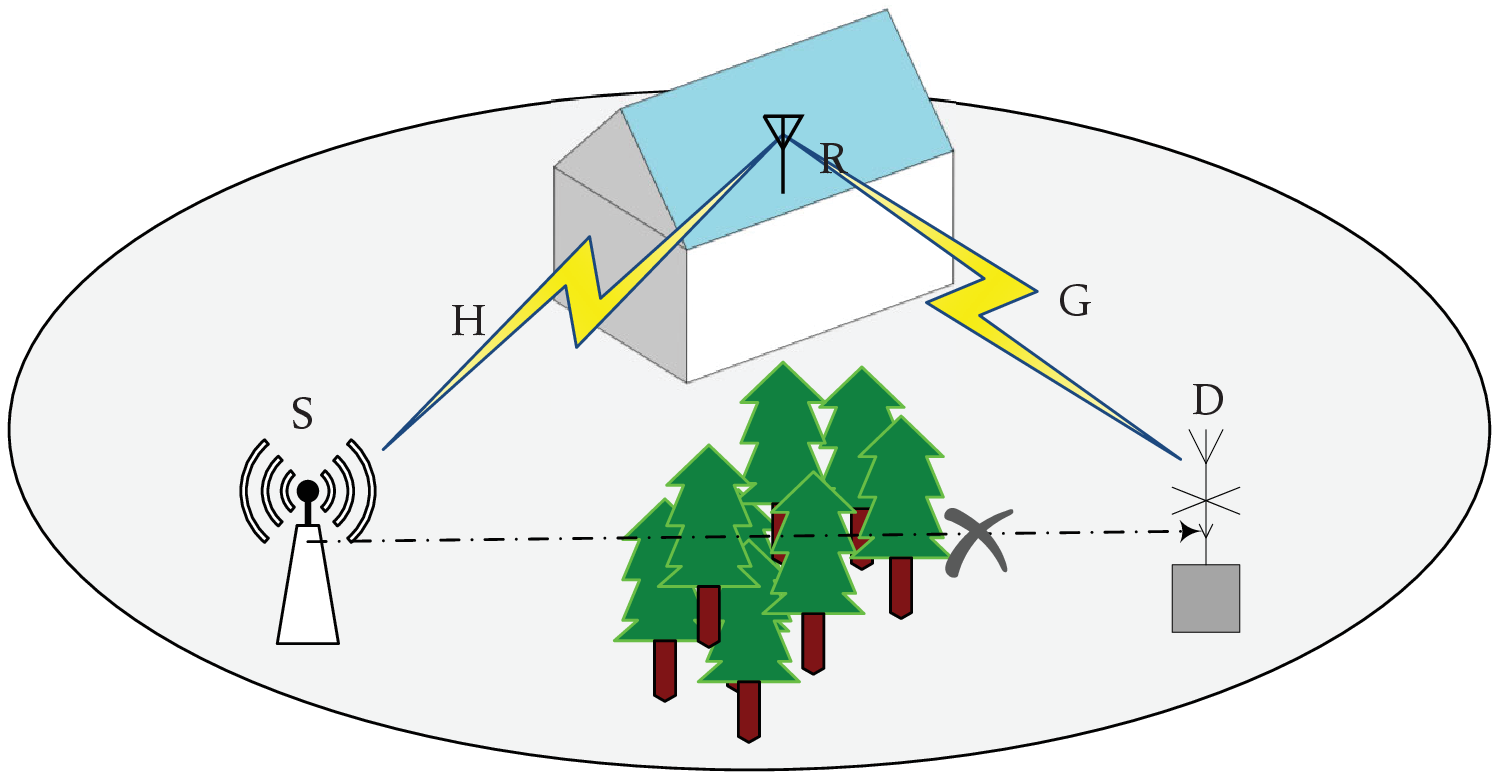}
	\caption{Relay supported transmission}
	\label{model2}
	\end{minipage}
	\end{figure}
	We focus on a single cell of a mmWave  wireless communication system and analyze two scenarios: when RIS is adopted and when AF relay is used, as shown in Fig. \ref{model1} and Fig. \ref{model2} respectively. {\color{black}Both the BS and user are assumed to be equipped with one antenna \footnote{{\color{blue}In general, large arrays of antennas are required in mmWave 5G systems to overcome the path-loss issue. Notice that our analysis can be generalized to the case of multiple-antenna BS or user easily, because every single FTR distribution in our model can be regarded as an approximation to the distribution of the sum of FTR RVs with the help of the accurate approximation method \cite{zhang2020Performance}. Thus, when the BS or user is equipped with multiple antennas, the element of channel coefficients of RIS-aided mmWave communications can still be modeled by a single FTR distribution.}}}. More specifically, in Fig. \ref{model1}, the RIS is set up between the BS and the user, comprising $L$ reflector elements arranged in a uniform array. In addition, the reflector elements are configurable and programmable via an RIS controller.
	In Fig. \ref{model2}, a half-duplex AF relay is deployed at the same location as the RIS.\footnote{\color{black}The reason why we consider an half-duplex AF relaying strategy is that DF introduces large complexity to the system \cite{azari2017ultra} and full-duplex technology can suffer considerable impediments due to the problem of self-interference. Moreover, when the FTR model is used in mmWave communications, the complexity in hardware impairments case of AF is lower than DF, and it will be easier to obtain insightful results. Furthermore, the RIS-aided system has been compared with the classic DF relaying in deterministic flat-fading channel where all the channel gains have the same squared magnitude \cite{bjornson2019intelligent}.} We consider the classic repetition coded AF relaying protocol \cite{J:Bjornsonetal} where the transmission is divided into two equal phases.
	
	More specifically, we assume that the direct transmission link between the BS and the user is blocked by trees or buildings. Thus, the RIS or relay is deployed to leverage the LoS paths to enhance the quality of received signals.  In addition, the fluctuating two-ray (FTR) fading model has been proposed in \cite{romero2017fluctuating} as a generalization of the two-wave with diffuse power fading model. {\color{black}The FTR fading model allows the constant amplitude specular waves of LoS propagation to fluctuate randomly, and incorporates ground reflections to provide a much better fit for small-scale fading measurements in mmWave communications \cite{zhang2017new}. In the following, we use the FTR fading model to illustrate the channel coefficients between the BS and the RIS, and the channel coefficients between the RIS and the user. Furthermore, an efficient channel state information (CSI) acquisition method for RIS-aided mmWave networks has been proposed \cite{cui2019efficient}, thus, we can assume that the parameters of FTR fading model is known to characterize the performance of the considered mmWave communication system.} {\color{black}Different from convenient microwave networks, the signal in mmWave communications is highly directional. Therefore, the high directivity implies that the interference among mmWave links can be significantly small. Thus, we ignore the interference from other cell-users.}
	
	\subsection{Exact Statistics of the End-to-End SNR For RIS-aided System}\label{SFSGA}
	{\color{black}To obtain the exact PDF and CDF of the end-to-end SNR for RIS-aided system, we first derive some important statistical expressions of FTR RVs.}
	\subsubsection{Exact Statistics of the Product of FTR RVs}
	The PDF and CDF of the squared FTR RV $\gamma$ are given respectively as \cite{zhang2017new}
	\begin{equation}\label{PDFFTR}
	{f_\gamma }\left( \gamma \right) = \frac{{{m^m}}}{{\Gamma (m)}}\sum\limits_{j = 0}^\infty  {\frac{{{K^j}{d_j}}}{{j!}}} {f_G}\left( {\gamma;j + 1,2{\sigma ^2}} \right),
	\end{equation}	
	\begin{equation}\label{CDFFTR}
	{F_\gamma }\left( \gamma \right) = \frac{{{m^m}}}{{\Gamma (m)}}\sum\limits_{j = 0}^\infty  {\frac{{{K^j}{d_j}}}{{j!}}} {F_G}\left( {\gamma;j + 1,2{\sigma ^2}} \right)
	\end{equation}	
	where
	\begin{equation}
	{f_G}\left( {\gamma;j + 1,2{\sigma ^2}} \right) \buildrel \Delta \over = \frac{{{\gamma^j}}}{{\Gamma (j + 1){{\left( {2{\sigma ^2}} \right)}^{j + 1}}}}\exp \left( { - \frac{x}{{2{\sigma ^2}}}} \right),
	\end{equation}
	\begin{equation}
	{F_G}\left( {x;j + 1,2{\sigma ^2}} \right) \buildrel \Delta \over = \frac{1}{{\Gamma (j + 1)}}\gamma \left( {j + 1,\frac{x}{{2{\sigma ^2}}}} \right),
	\end{equation}
	and
	\begin{align}
	{d_n} \buildrel \Delta \over =& \sum\limits_{k = 0}^n {\left( {\begin{array}{*{20}{l}}
	n\\
	k
	\end{array}} \right)} {\left( {\frac{\Delta }{2}} \right)^k}\sum\limits_{l = 0}^k {\left( {\begin{array}{*{20}{c}}
	k\\
	l
	\end{array}} \right)} \Gamma (n + m + 2l - k){e^{\frac{{\pi (2l - k)j}}{2}}}\notag\\
	&\times {\left( {{{(m + K)}^2} - {{(K\Delta )}^2}} \right)^{\frac{{ - (n + m)}}{2}}}P_{n + m - 1}^{k - 2l}\left( {\frac{{m + K}}{{\sqrt {{{(m + K)}^2} - {{(K\Delta )}^2}} }}} \right)
	\end{align}
	
	Substituting $R = \sqrt \gamma $ into \eqref{PDFFTR}, we can easily obtain the PDF expression of FTR RVs. {\color{black}Because the channel coefficients between the BS and the RIS and between the RIS and the user are modeled by FTR distribution, for one element of RIS, the product channel between the BS and the user should be modeled by the distribution of the product of FTR RVs. Thus, let $X \buildrel \Delta \over = \prod\limits_{\ell  = 1}^N {{R_\ell }}$ as the product of $N$ FTR RVs, where $R \sim {\cal F}{\cal T}{\cal R}\left( {{K_\ell },{m_\ell },{\Delta _\ell },\sigma _\ell ^2} \right)$ ($\ell=1,\cdots,N$), we can derive the exact $s_{\rm th}$ moment, PDF and CDF expressions for $X$, which are summarized in the following Theorems.}
	\begin{them}\label{P1}
	The $s{\rm th}$ moment of $X$ is given by
	\begin{equation}\label{expdex}	
	{\mathbb E}\left[ {{X^s}} \right] = \prod\limits_{\ell  = 1}^N {\frac{{{m_\ell }^{{m_\ell }}}}{{\Gamma ({m_\ell })}}\sum\limits_{{j_\ell } = 0}^\infty  {\frac{{{K_\ell }^{{j_\ell }}{d_\ell }_{{j_\ell }}}}{{{j_\ell }!}}} \frac{{{{\left( {2{\sigma _\ell }^2} \right)}^{^{\frac{s}{2}}}}}}{{\Gamma ({j_\ell } + 1)}}} \Gamma \left( {1 + {j_\ell } + \frac{s}{2}} \right).
	\end{equation}
	\end{them}
	\begin{IEEEproof}
	The $s{\rm th}$ order moment of $X$ is derived as	
	\begin{equation}\label{moment}
	{\mathbb E}\left[ {{X^s}} \right] = \prod\limits_{\ell  = 1}^N {\mathbb E} \left[ {R_\ell ^s} \right] = \prod\limits_{\ell  = 1}^N {\int_0^\infty  {{r^s}} } {f_{{R_\ell }}}(r)dr.
	\end{equation}
	By substituting \eqref{PDFFTR} into \eqref{moment} and using \cite[eq. (3.381.10)]{gradshteyn2007} and \cite[eq. (8.356.3)]{gradshteyn2007}, we obtain the $s{\rm th}$ moment of $X$, which completes the proof.	
	\end{IEEEproof}
	\begin{them}\label{APPAT}
	The PDF, CDF and MGF expressions of $X$ can be deduced in closed-form as	
	{\small \begin{equation}\label{PDFFINAL1}
	{f_X}(x) =\!\! \sum\limits_{{j_1,\cdots,j_N } = 0}^\infty  {\prod\limits_{\ell  = 1}^N {\frac{{{m_\ell }^{{m_\ell }}}}{{\Gamma ({m_\ell })}}\frac{{2{x^{ - 1}}}}{{\Gamma ({j_\ell } + 1)}}} \frac{{{K_\ell }^{{j_\ell }}{d_\ell }_{{j_\ell }}}}{{{j_\ell }!}}G_{0,N}^{N,0}\left( {\left. {{x^2}\prod\limits_{\ell  = 1}^N {\left( {\frac{1}{{2{\sigma _\ell }^2}}} \right)} } \right|\begin{array}{*{20}{c}}
	- \\
	{1 + {j_1}, \cdots ,1 + {j_N}}
	\end{array}} \right)},
	\end{equation}
	\begin{equation}\label{CDFPRO}
	{F_X}\left( x \right) =\!\! \sum\limits_{{j_1,\cdots,j_N } = 0}^\infty  {\prod\limits_{\ell  = 1}^N {\frac{{{m_\ell }^{{m_\ell }}}}{{\Gamma ({m_\ell })}}\frac{1}{{\Gamma ({j_\ell } + 1)}}\frac{{{K_\ell }^{{j_\ell }}{d_\ell }_{{j_\ell }}}}{{{j_\ell }!}}G_{1,N + 1}^{N,1}\left( {\left. {{x^{\rm{2}}}\prod\limits_{\ell  = 1}^N {\left( {\frac{1}{{2{\sigma _\ell }^2}}} \right)} } \right|\begin{array}{*{20}{c}}
	1\\
	{1 + {j_1}, \cdots ,1 + {j_N},0}
	\end{array}} \right)} } ,
	\end{equation}
	\begin{equation}\label{MGFPRO}
	{{\cal M}_X}\left( s \right) = \!\!\sum\limits_{{j_1,\cdots,j_N} = 0}^\infty  {\prod\limits_{\ell  = 1}^N {\frac{{{K_\ell }^{{j_\ell }}{d_\ell }_{{j_\ell }}}}{{{j_\ell }!}}\frac{{{m_\ell }^{{m_\ell }}}}{{\Gamma ({m_\ell })}}\frac{1}{{\Gamma ({j_\ell } + 1)}}} } H_{1,1}^{1,1}\left( {\left. {\frac{1}{{s\prod\limits_{\ell  = 1}^N {\left( {\sqrt 2 {\sigma _\ell }} \right)} }}} \right|\begin{array}{*{20}{c}}
	{\left( {1,1} \right)}\\
	{\left( {1 + {j_\ell },0.5} \right)}
	\end{array}} \right)
	\end{equation}}
	where $G \, \substack{ m , n \\ p , q}(\cdot)$ is the Meijer's $G$-function \cite[eq. (9.301)]{gradshteyn2007} and $H \, \substack{ m , n \\ p , q}(\cdot)$ is the Fox's  $H$-function \cite[eq. (1.2)]{mathai2009h}.
	\end{them}
	\begin{IEEEproof}Please refer to Appendix \ref{AppendixA}.\end{IEEEproof}
	\begin{rem}
	Although the PDF, CDF and MGF of a product of $N$ i.n.i.d. squared FTR RVs have been respectively derived in \cite[eq. (7)]{badarneh2019cascaded}, \cite[eq. (16)]{badarneh2019cascaded} and \cite[eq. (17)]{badarneh2019cascaded}. Only the parameter $m$ of each FTR RV is different from each other. In the practical mmWave communication scenario, all parameters $K,m,\Delta $ and $\sigma ^2$ of each FTR RV can be different, thus the statistical characterizations obtained in Theorem \ref{APPAT} are more general and useful in the performance analysis.
	\end{rem}
	\begin{rem}
	Substituting $\gamma = x^2$ into \eqref{PDFFINAL1} and \eqref{CDFPRO}, we can obtain more general statistical expressions of the product of an arbitrary number of i.n.i.d. squared FTR RVs. Besides, note that the bivariate Meijer's $G$-function can be evaluated numerically in an efficient manner using the MATLAB program \cite{J:Peppas2012WCL}, and two Mathematica implementations of the single Fox's $H$-function are provided in \cite{sanchez2019statistics} and \cite{8703806}.
	\end{rem}
	\subsubsection{Exact PDF and CDF of the Sum of Product of FTR RVs}\label{sumpro}
	{\color{black}Because RIS has a large number of elements, the channel between the BS and the user should be modeled by the distribution of the sum of product of FTR RVs. In this subsection, we derive the exact PDF and CDF of the sum of product of FTR RVs in terms of multivariate Fox's $H$-function \cite[eq. (A-1)]{mathai2009h}.}
	\begin{them}\label{sumrvftr}
	We define $ Y = \sum\limits_{\iota=1} ^L {{X_\iota }}  $. Thus, the PDF and CDF of $Y$ can be deduced in closed-form as
	{\small \begin{align}\label{PDFFTRSUMPRO}
	&{f_Y}(y){=}\sum\limits_{{j_{1.1}}, \cdots ,{j_{L,1}} = 0}^\infty   \cdots  \sum\limits_{{j_{1.N}}, \cdots ,{j_{L,N}} = 0}^\infty  {\prod\limits_{\iota  = 1}^L {\prod\limits_{\ell  = 1}^N {\frac{{{K_{\iota ,\ell }}^{{j_{\iota ,\ell }}}{d_{\iota ,\ell }}_{{j_{\iota ,\ell }}}}}{{{j_{\iota ,\ell }}!}}\frac{{{m_{\iota ,\ell }}^{{m_{\iota ,\ell }}}}}{{\Gamma ({m_{\iota ,\ell }})}}\frac{1}{{\Gamma ({j_{\iota ,\ell }} + 1)}}\frac{1}{y}} } }
	\notag\\&\times
	H_{{{1,0:N,1;}} \cdots {{;N,1}}}^{{{0,0:1,N;}} \cdots {{;1,N}}}\left( {\left. {\begin{array}{*{20}{c}}
	{{y^{ - 1}}\prod\limits_{\ell  = 1}^N {\left( {\sqrt 2 {\sigma _{{{1}},\ell }}} \right)} }\\
	\vdots \\
	{{y^{ - 1}}\prod\limits_{\ell  = 1}^N {\left( {\sqrt 2 {\sigma _{L,\ell }}} \right)} }
	\end{array}} \right|\begin{array}{*{20}{c}}
	{\left( {{{0;1,}} \cdots ,{{1}}} \right):\left\{ {\left( { - {j_{1,n}},0.5} \right)} \right\}_1^N; \cdots ;\left\{ {\left( { - {j_{L,n}},0.5} \right)} \right\}_1^N}\\
	{ - :\left( {0,1} \right); \cdots ;\left( {0,1} \right)}
	\end{array}} \right),
	\end{align}}
	\vspace{-5mm}
	{\small \begin{align}\label{CDFFTRSUMPRO}
	{F_Y}(y){=}&\sum\limits_{{j_{1.1}}, \cdots ,{j_{L,1}} = 0}^\infty   \cdots  \sum\limits_{{j_{1.N}}, \cdots ,{j_{L,N}} = 0}^\infty  {\prod\limits_{\iota  = 1}^L {\prod\limits_{\ell  = 1}^N {\frac{{{K_{\iota ,\ell }}^{{j_{\iota ,\ell }}}{d_{\iota ,\ell }}_{{j_{\iota ,\ell }}}}}{{{j_{\iota ,\ell }}!}}\frac{{{m_{\iota ,\ell }}^{{m_{\iota ,\ell }}}}}{{\Gamma ({m_{\iota ,\ell }})}}\frac{1}{{\Gamma ({j_{\iota ,\ell }} + 1)}}} } }\notag\\&\times H_{{{1,0:N,1;}} \cdots {{;N,1}}}^{{{0,0:1,N;}} \cdots {{;1,N}}}\left( {\left. {\begin{array}{*{20}{c}}
	{{y^{ - 1}}\prod\limits_{\ell  = 1}^N {\left( {\sqrt 2 {\sigma _{{{1}},\ell }}} \right)} }\\
	\vdots \\
	{{y^{ - 1}}\prod\limits_{\ell  = 1}^N {\left( {\sqrt 2 {\sigma _{L,\ell }}} \right)} }
	\end{array}} \right|\begin{array}{*{20}{c}}
	{\left( {{{1;1,}} \cdots ,{{1}}} \right):\left\{ {\left( { - {j_{1,n}},0.5} \right)} \right\}_1^N; \cdots ;\left\{ {\left( { - {j_{L,n}},0.5} \right)} \right\}_1^N}\\
	{ - :\left( {0,1} \right); \cdots ;\left( {0,1} \right)}
	\end{array}} \right)
	\end{align}}
	where $ \left\{ {\left( {{a_n}} \right)} \right\}_1^N = \left( {{a_1}} \right), \cdots ,\left( {{a_N}} \right) $.
	\end{them}
	\begin{IEEEproof}Please refer to Appendix \ref{AB}.\end{IEEEproof}
	\begin{rem}
	Although the numerical evaluation for multivariate Fox's $H$-function is unavailable in popular mathematical packages such as MATLAB and Mathematica, its efficient implementations have been reported in recent literature.. For example, a Python implementation for the multivariable Fox's $H$-function is presented in \cite{alhennawi2015closed}, and an efficient GPU-oriented MATLAB routine for the multivariate Fox's $H$-function is introduced in \cite{chergui2018rician}. In the following, we will utilize these novel implementations to evaluate our results.
	\end{rem}
	\subsubsection{SNR Analysis}
	{\color{black}
	We focus on the downlink of the RIS-aided system. The signal received from the BS through the RIS for the user is given by \cite[eq. (10)]{basar2019wireless}
	\begin{equation}
	y = \sqrt P {\bf g}{\bf \Phi} {\bf h}s + n
	\end{equation}	
	where $ {\bf \Phi}  = \beta{\rm  diag} \left[ {{e^{i{\phi _1}}}, \cdots ,{e^{i{\phi _L}}}} \right] $, $ \beta  \in \left( {{{0}},{{1}}} \right] $ is the fixed amplitude reflection coefficient\footnote{In practice, each element of the RIS is usually designed to maximize the signal reflection. Thus, we set $\beta = 1$ for simplicity.} and $ {\phi _1}, \cdots ,{\phi _L} $ are the phase shifts that can be optimized by the RIS controller. {\color{blue}The channel coefficients between the BS and the RIS are denoted by an $L \times 1$ vector $ {\bf h}$, and the channel coefficients between the RIS and the user can be expressed as an $1 \times L$ vector $ {\bf g}$, where the elements of ${\bf h}$ and ${\bf g}$ are i.n.i.d. FTR RVs.}
	
	Notice that the elements of ${\bf h}$ or ${\bf g}$ will have different phases, because under the FTR model, the complex baseband voltage of a wireless channel experiencing multipath fading contains two fluctuating specular components with different phases. Moreover, the angle of departure (AoD) and angle of arrival (AoA) of a signal will also cause phase difference \cite{han2019large}. Thus, ${\bf h}$ can be written as \cite[Section \uppercase\expandafter{\romannumeral 4}]{basar2019wireless}
	\begin{equation}\label{hzheli}
	{\bf h} = {\left[ {{h_1}{e^{i{\theta _{1,1}}}}, \cdots ,{h_L}{e^{i{\theta _{L,1}}}}} \right]^{\rm H}}
	\end{equation}
	where $h_\ell$ denotes the amplitude of BS-RIS channel coefficient, and ${{\theta _{\ell,1}}}$ is the corresponding phase.
	
	Similarly, $ {\bf g}$ can be expressed as \cite[Section \uppercase\expandafter{\romannumeral 4}]{basar2019wireless}
	\begin{equation}\label{gzheli}
	{\bf g} = \left[ {{g_1}{e^{i{\theta _{1,{{2}}}}}}, \cdots ,{g_L}{e^{i{\theta _{L,2}}}}} \right]
	\end{equation}
	where $g_\ell$ denotes the amplitude of RIS-user channel coefficient, and ${{\theta _{\ell,2}}}$ is the corresponding phase.}
	
	Accordingly, the SNR of RIS-aided system is given by \cite[eq. (11)]{basar2019wireless}	
	\begin{equation}\label{snrris}
	{\gamma _{{\rm RIS}}'} = {\left| {\sum\limits_{\ell  = 1}^L {{h_\ell }{g_\ell }{e^{i({\theta _{\ell ,1}} + {\theta _{\ell ,2}} + {\phi _\ell })}}} } \right|^2}\frac{P}{{{o ^{{2}}}}}.
	\end{equation}
	Before designing the phase shift, we can theoretically obtain the maximum of ${\gamma _{{\rm RIS}}'}$ with the optimal phase shift design of RIS's reflector array as
	\begin{equation}\label{maxris}
	{\gamma _{{\rm RIS}}} = \frac{P}{{{o ^{{2}}}}}{\left( {\sum\limits_{\ell  = 1}^L {{h_\ell }{g_\ell }} } \right)^2}.
	\end{equation}
	{\color{black}Because the electrical size of the unit cell of programmable metasurface is between $\lambda/8$ and $\lambda/4$ in principle, where $\lambda$ is a wavelength of the signal \cite{scui2017information}, channel correlation may exist between the RIS-related channels. However, there is no experiments or field trials have been reported to model the correlation at the RIS. Therefore, we consider the weak correlation scenario, i.e., the BS communicates with the user using RIS in the far-field region \cite{dardari2019communicating,bjornson2020power}. Thus, we assume that the small scale fading components of the channel, ${h_\ell }$ and ${g_\ell }$, are independent of each other.} {\color{black}Thus, with the help of the statistic of the sum of product of FTR RVs derived in Section \ref{sumpro}, we obtain the following corollary.}
	\begin{corr}
	The PDF and CDF of the end-to-end SNR for RIS-aided system, ${\gamma _{RIS}}$, can be derived as
	{\small \begin{align}\label{cor1pdf}
	{f_{{\gamma _{RIS}}}}(z){{ = }}&\frac{{{1}}}{{2z}}\sum\limits_{{j_{1.1}}, \cdots ,{j_{L,1}} = 0}^\infty  {\sum\limits_{{j_{1.N}}, \cdots ,{j_{L,{{2}}}} = 0}^\infty  {\prod\limits_{\iota  = 1}^L {\prod\limits_{\ell  = 1}^2 {\frac{{{K_{\iota ,\ell }}^{{j_{\iota ,\ell }}}{d_{\iota ,\ell }}_{{j_{\iota ,\ell }}}}}{{{j_{\iota ,\ell }}!}}\frac{{{m_{\iota ,\ell }}^{{m_{\iota ,\ell }}}}}{{\Gamma ({m_{\iota ,\ell }})}}\frac{1}{{\Gamma ({j_{\iota ,\ell }} + 1)}}} } } }
	\notag\\&\times
	H_{{{1,0:2,1;}} \cdots {{;2,1}}}^{{{0,0:1,2;}} \cdots {{;1,2}}}\left(\!\! {\left. {\begin{array}{*{20}{c}}
	{{{\left( {\frac{P}{{{o^2}z}}} \right)}^{\frac{1}{2}}}\prod\limits_{\ell  = 1}^2 {\left( {\sqrt 2 {\sigma _{{{1}},\ell }}} \right)} }\\
	\vdots \\
	{{{\left( {\frac{P}{{{o^2}z}}} \right)}^{\frac{1}{2}}}\prod\limits_{\ell  = 1}^2 {\left( {\sqrt 2 {\sigma _{L,\ell }}} \right)} }
	\end{array}} \!\!\right|\!\!\begin{array}{*{20}{c}}
	{\left( {{{0;1,}} \cdots ,{{1}}} \right):\left\{ {\left( { - {j_{1,n}},0.5} \right)} \right\}_1^2; \cdots ;\left\{ {\left( { - {j_{L,n}},0.5} \right)} \right\}_1^2}\\
	{ - :\left( {0,1} \right); \cdots ;\left( {0,1} \right)}
	\end{array}} \!\!\right),
	\end{align}}
	{\small\begin{align}\label{cor1cdf}
	{F_{{\gamma _{RIS}}}}(z){{ = \!\!}}&\sum\limits_{{j_{1.1}}, \cdots ,{j_{L,1}} = 0}^\infty  {\sum\limits_{{j_{1.N}}, \cdots ,{j_{L,{{2}}}} = 0}^\infty  {\prod\limits_{\iota  = 1}^L {\prod\limits_{\ell  = 1}^2 {\frac{{{K_{\iota ,\ell }}^{{j_{\iota ,\ell }}}{d_{\iota ,\ell }}_{{j_{\iota ,\ell }}}}}{{{j_{\iota ,\ell }}!}}\frac{{{m_{\iota ,\ell }}^{{m_{\iota ,\ell }}}}}{{\Gamma ({m_{\iota ,\ell }})}}\frac{1}{{\Gamma ({j_{\iota ,\ell }} + 1)}}} } } }
	\notag\\&\times
	H_{{{1,0:2,1;}} \cdots {{;2,1}}}^{{{0,0:1,2;}} \cdots {{;1,2}}}\left( \!\!\!{\left. {\begin{array}{*{20}{c}}
	{{{\left( {\frac{P}{{{o^2}z}}} \right)}^{\frac{1}{2}}}\prod\limits_{\ell  = 1}^2 {\left( {\sqrt 2 {\sigma _{{{1}},\ell }}} \right)} }\\
	\vdots \\
	{{{\left( {\frac{P}{{{o^2}z}}} \right)}^{\frac{1}{2}}}\prod\limits_{\ell  = 1}^2 {\left( {\sqrt 2 {\sigma _{L,\ell }}} \right)} }
	\end{array}} \!\right|\!\!\begin{array}{*{20}{c}}
	{\left( {{{1;1,}} \cdots ,{{1}}} \right):\left\{ {\left( { - {j_{1,n}},0.5} \right)} \right\}_1^2; \cdots ;\left\{ {\left( { - {j_{L,n}},0.5} \right)} \right\}_1^2}\\
	{ - :\left( {0,1} \right); \cdots ;\left( {0,1} \right)}
	\end{array}} \!\!\!\right).
	\end{align}}
	\end{corr}
	
	\begin{IEEEproof}
	Using Theorem \ref{sumrvftr} and letting $N=2$, we obtain the PDF and CDF of the sum of product of two FTR RVs.  With the help of \eqref{maxris}, eqs. \eqref{cor1pdf} and \eqref{cor1cdf} can be derived after some transformation of RVs, which completes the proof.
	\end{IEEEproof}
	{\color{black}The statistic of ${\gamma _{RIS}}$ derived in this subsection can be used to obtain exact closed-form expressions for key performance metrics of the RIS-aided mmWave communications, such as OP and ABEP.}
	
	\subsection{Exact Statistics of the End-to-End SNR For AF Relay System}
	We consider the classic AF relay protocol where the transmission is divided into two equal-sized phases. The transmit power of BS is $P_1$ in the first phase, and one of AF relay is $P_2$ in the second phase. {\color{black}Assuming variable gain relays, the end-to-end SNR of AF relay system is given as \cite{J:Bjornsonetal}
	\begin{equation}\label{Eq:SNRhard1}
	{\gamma _{{\rm{AF}}}'}  = \frac{\gamma_1 \, \gamma_2}{d_h \, \gamma_1 \, \gamma_2 +c_{h,1} \, \gamma_1 +c_{h,2} \, \gamma_2 + 1}
	\end{equation}
{\color{blue}where $c_{h,i} = 1+\kappa_i^2$, $d_h \triangleq \kappa_1^2 \, \kappa_2^2+{\kappa_1}^2+{\kappa_2}^2$, and $\kappa_1$ and $\kappa_2$ characterize the level of impairments in the BS and relay hardware, respectively.}
	Eq. \eqref{Eq:SNRhard1} can be tightly approximated as \cite[eq. (9)]{karatza2019unified}
	\begin{equation}\label{Eq:SNRhard2}
	{\gamma_\text{AF}} \approx  \frac{\gamma_1 \, \gamma_2}{d_h \, \gamma_1 \, \gamma_2 +c_{h,1} \, \gamma_1 +c_{h,2} \, \gamma_2 } .
	\end{equation}	
	This approximation has been used in several works to analyze the performance of AF relay system, e.g. \cite{hasna2003outage,karatza2019unified,zhang2020dual}, and it is tight at medium and high SNR values. Furthermore, eq. \eqref{Eq:SNRhard2} can be regarded as the upper bound of $\gamma_{{\rm{AF}}}'$ in low SNR regimes. Thus, if we figure out how a RIS-aided system can outperform the AF relay system which has a slightly higher SNR, we can still find the answer of the question \textit{``How can a RIS outperform AF relaying over realistic mmWave channels?"}.  }
	For ideal hardware, i.e. $c_{h,i} = 1$ and $d_h = 0$, eq. \eqref{Eq:SNRhard2} reduces to the half-harmonic mean of $\gamma_\ell$ as 
	\begin{equation}\label{GAMMAAF}
	{\gamma _\text{AF}} \approx \frac{{{\gamma _1}{\gamma _2}}}{{{\gamma _1} + {\gamma _2}}}.
	\end{equation}
	{\color{blue}Assuming that $ {\gamma _i}{=}\frac{{{P_i}}}{{{o ^{{2}}}}}{\left| {{q_i}} \right|^{{2}}} \left( {i = 1,2} \right) $, and $Z\triangleq\frac{{{{\left| {{q_{{1}}}} \right|}^{{2}}}{{\left| {{q_2}} \right|}^{{2}}}}}{{d{{\left| {{q_{{1}}}} \right|}^{{2}}}{{\left| {{q_2}} \right|}^{{2}}} + {c_1}{{\left| {{q_{{1}}}} \right|}^{{2}}} + {c_2}{{\left| {{q_2}} \right|}^{{2}}}}}$, we can rewrite \eqref{Eq:SNRhard2} as
		{\small \begin{equation}\label{AFDESNR}
			{\gamma_\text{AF}} \approx \frac{{{P_{{1}}}{P_2}}}{{{o ^{{2}}}}}Z
			\end{equation}}}
	where $d=P_1 P_2 d_h$ and $c_{i}=P_{i} c_{h,i},\left( {i = 1,2} \right) $.
	
	An integral representation for the PDF of ${\gamma_\text{AF}}$ in \eqref{Eq:SNRhard2} assuming arbitrarily distributed $\gamma_\ell$ and considering hardware impairments is given by \cite[eq. (16)]{zhang2020dual}
{\small 	\begin{equation}\label{jifenafpdf}
		{f_{{{\gamma_\text{AF}}}}}\left( \gamma \right) = \frac{{{c_1}{c_2}\gamma}}{{{{(1 - \gamma d)}^3}}}\int_0^1 {{f_{{\gamma _1}}}} \left( {\frac{{{c_2}\gamma}}{{(1 - \gamma d)t}}} \right){f_{{\gamma _2}}}\left( {\frac{{{c_1}\gamma}}{{(1 - \gamma d)(1 - t)}}} \right)\frac{{{\rm{d}}t}}{{{t^2}{{(1 - t)}^2}}}.
\end{equation}}
	To derive a generic analytical expression for the CDF of the end-to-end SNR of AF relay systems with hardware impairments, we first present the following useful Lemma.
	\begin{lemma}\label{ACL}
	Assuming arbitrarily distributed $\gamma$ and considering hardware impairments, an integral representation for the CDF of $\gamma_{AF}$ in \eqref{Eq:SNRhard2}  is given by
	{\small \begin{align}\label{zhuizhongCDF}
			{F_{{\gamma_\text{AF}}}}\left( \gamma \right) = {1 - \frac{{\gamma{c_1}}}{{\left( {1 - \gamma d} \right)}}\int_0^1 {{f_{{\gamma _2}}}\left( {\frac{{\gamma {c_1}}}{{\left( {1 - \gamma d} \right)t}}} \right)} \frac{{{\rm d}t}}{{{t^2}}}}
			+ \frac{{\gamma{c_1}}}{{\left( {1 - \gamma d} \right)}} \int_0^1 {{F_{{\gamma _1}}}\left( {\frac{{\gamma{c_2}}}{{\left( {1 - \gamma d} \right)\left( {1 - t} \right)}}} \right)} {f_{{\gamma _2}}}\left( {\frac{{\gamma{c_1}}}{{\left( {1 - \gamma d} \right)t}}} \right)\frac{{{\rm d}t}}{{{t^2}}} .
	\end{align}}
	\end{lemma}
	\begin{IEEEproof}	
	Please refer to Appendix \ref{AC}.
	\end{IEEEproof}
	{\color{black}Notice that the integral representation for the CDF of AF relay system obtained in Lemma \ref{ACL} can be used to analyze arbitrarily distributed $\gamma_{\rm{AF}}$. For the special case of FTR-distributed hops, with the help of integral representations for the PDF and CDF, \eqref{jifenafpdf} and \eqref{zhuizhongCDF}, closed-form statistics of AF relay system can be derived for both cases of non-ideal and ideal hardware.}
	\begin{them}\label{them4}	
	Considering hardware impairments, the PDF and CDF of $Z$ can be deduced in closed-form as
	{\small \begin{align}	\label{PDFAF}
	{f_{{Z}}}(z) \!=\! \frac{{{c_2}{c_1}z}}{{{{(1\! -\! zd)}^3}}}\!\sum\limits_{{j_{{1}}} = 0}^\infty  {\sum\limits_{{j_{{2}}} = 0}^\infty  {\prod\limits_{\ell  = 1}^2 { {{\frac{{{m_\ell }^{{m_\ell }}K_\ell ^{{j_\ell }}{d_{{\ell _{{j_\ell }}}}}A_\ell ^{{j_\ell }}}}{{2\sigma _\ell ^2\Gamma \left( {{j_\ell } + 1} \right)\Gamma \left( {{m_\ell }} \right){j_\ell }!}}}} } } } H_{0,1:2,0;2,0}^{0,0:0,2;0,2}\left(\!\!\!\! {\left. {\begin{array}{*{20}{c}}
	{A_1^{ - 1}}\\
	{A_2^{ - 1}}
	\end{array}} \!\!\right|\begin{array}{*{20}{c}}\!
	{ - :\left( {1,1} \right)\left( {2\! +\! {j_1},1} \right)\!;\!\left( {1,1} \right)\left( {2 + {j_2},1} \right)}\\
	{\left( {2 + {j_1} + {j_2};1,1} \right): - }
	\end{array}}\!\! \!\!\right),
	\end{align}}
	{\small \begin{equation}\label{CDFAF}
	{F_{{Z}}}\left( z \right) ={1 - { \frac{{{m_{2}}^{{m_{{2}}}}}}{{\Gamma ({m_{2}})}}\sum\limits_{{j_{2}} = 0}^\infty  {\frac{{{K_{2}}^{{j_{{2}}}}{d_{2}}_{{j_{2}}}}}{{{j_{2}}!}}} \frac{1}{{\Gamma ({j_{2}} + 1)}}\Gamma \left( {1 + {j_2},{A_2}} \right)}}  { + {{H_{AF}}}}
	\end{equation}}
	where $ {A_1} = \frac{{{c_2}z}}{{2{\sigma _{{1}}}^2\left( {1 - zd} \right)}} $, $ {A_2} = \frac{{{c_1}z}}{{2{\sigma _2}^2\left( {1 - zd} \right)}} $ and
	{\small \begin{align}
	{H_{AF}}\! = \!\frac{{z{c_1}}}{{\left( {1 \!- \!zd} \right)}}\sum\limits_{{j_{1}} = 0}^\infty  {\sum\limits_{{j_{2}} = 0}^\infty  {\prod\limits_{\ell  = 1}^2 {{\frac{{{\left( {{A_2}} \right)}^{{j_{2}}}}{{{m_\ell }^{{m_\ell }}}{K_\ell }^{{j_\ell }}{d_\ell }_{{j_\ell }}}}{{2{\sigma _{{2}}}^2}{ \Gamma ({j_\ell } + 1)}{\Gamma ({m_\ell }){j_\ell }!}}} } } } H_{{0},{{1:3,1;2,0}}}^{{{0}},{{0:1,2;0,2}}}\left(\!\!\!\! {\left. {\begin{array}{*{20}{c}}
	{{A_1}^{ - 1}}\\
	{{A_2}^{ - 1}}
	\end{array}} \!\!\right|\!\!\begin{array}{*{20}{c}}
	{\! -\! :\!\left( {0,1} \right)\left( {-j_1,1} \right)\left( {1,1} \right)\!;\!\left( {1,1} \right)\left( {2 + {j_2},1} \right)}\\
	{\left( {1 + {j_2};1,1} \right):\left( {0,1} \right); - }
	\end{array}} \!\!\!\!\right).
	\end{align}}
	\end{them}
	\begin{IEEEproof}
	Please refer to Appendix \ref{AT4}.
	\end{IEEEproof}	
	To compare the AF relay system with the RIS-aided system, we consider the ideal hardware to make a fair comparison.
	\begin{corr}	
	For the special case of ideal hardware, the PDF and CDF of ${\gamma_{{\rm{AF}}}}$ can be deduced in closed-form as
	{\small \begin{align}\label{PDFAFSVC}
	{f_{{\gamma_\text{AF}}}}\left( z \right) =&\frac{{{o^4}}}{{{P_1}{P_2}}}z\sum\limits_{{j_{{1}}} = 0}^\infty  {\sum\limits_{{j_{{2}}} = 0}^\infty  {\prod\limits_{\ell  = 1}^2 {\left( {\frac{{{m_\ell }^{{m_\ell }}{K_\ell }^{{j_\ell }}{d_\ell }_{{j_\ell }}}}{{2{\sigma _\ell }^2\Gamma \left( {{j_\ell } + 1} \right)\Gamma ({m_\ell }){j_\ell }!}}} \right)} } } {\left( {\frac{{{o^2}z}}{{{{2}}\sigma _1^2{P_{{1}}}}}} \right)^{{j_1}}}{\left( {\frac{{{o^2}z}}{{{{2}}\sigma _{{2}}^2{P_{{2}}}}}} \right)^{{j_2}}}
	\notag\\
	&\times H_{0,1:2,0;2,0}^{0,0:0,2;0,2}\left( {\left. {\begin{array}{*{20}{c}}
	{\frac{{{{2}}\sigma _1^2{P_{{1}}}}}{{{o^2}z}}}\\
	{\frac{{{{2}}\sigma _2^2{P_2}}}{{{o^2}z}}}
	\end{array}} \right|\begin{array}{*{20}{c}}
	{ - :\left( {1,1} \right)\left( {2 + {j_1},1} \right);\left( {1,1} \right)\left( {2 + {j_2},1} \right)}\\
	{\left( {2 + {j_1} + {j_2};1,1} \right): - }
	\end{array}} \right),
	\end{align}}
	\vspace{-5mm}
	{\small \begin{align}\label{CDFAFOUTH}
	{F_{{\gamma_\text{AF}}}}\left( z \right) \!=& 1 \!-\!\frac{{{m_{{2}}}^{{m_{{2}}}}}}{{\Gamma ({m_{{2}}})}}\sum\limits_{{j_{{2}}} = 0}^\infty  {\frac{{{K_{{2}}}^{{j_{{2}}}}{d_{{2}}}_{{j_{{2}}}}}}{{{j_{{2}}}!}}} \frac{1}{{\Gamma ({j_{{2}}} + 1)}}\Gamma \left( {1 + {j_2},\frac{{{o^2}z}}{{{{2}}\sigma _{{2}}^2{P_{{2}}}}}} \right) \! + \!\sum\limits_{{j_{{1}}} = 0}^\infty  {\sum\limits_{{j_{{2}}} = 0}^\infty  {\prod\limits_{\ell  = 1}^2 { {\frac{{{m_\ell }^{{m_\ell }}}}{{\Gamma ({m_\ell })}}\frac{{{K_\ell }^{{j_\ell }}{d_\ell }_{{j_\ell }}}}{{{j_\ell }!}}\frac{{{1}}}{{\Gamma ({j_\ell } + 1)}}} } } } {\left( {\frac{{{o^2}z}}{{{{2}}\sigma _{{2}}^2{P_{{2}}}}}} \right)^{{j_{{2}}} + 1}}	\notag\\
	&\times H_{{{0,1:3,2;2,0}}}^{{{0,0:1,2;0,2}}}\left( {\left. {\begin{array}{*{20}{c}}
	{\frac{{{{2}}\sigma _1^2{P_{{1}}}}}{{{o^2}z}}}\\
	{\frac{{{{2}}\sigma _2^2{P_2}}}{{{o^2}z}}}
	\end{array}} \right|\begin{array}{*{20}{c}}
	{ - :\left( {0,1} \right)\left( {-j_1,j_1} \right)\left( {1,1} \right);\left( {1,1} \right)\left( {2 + {j_2},1} \right)}\\
	{\left( {1 + {j_2};1,1} \right):\left( {0,1} \right); - }
	\end{array}} \right).	
	\end{align}}
	\end{corr}
	\begin{IEEEproof}
	Setting $c_1 =P_1$, $c_2 = P_2$ and $d = 0$ in \eqref{PDFAF} and \eqref{CDFAF}, we can obtain \eqref{PDFAFSVC} and \eqref{CDFAFOUTH} to complete the proof.
	\end{IEEEproof}
	{\color{black}The statistic of ${{\gamma_\text{AF}}}$ we obtained in this subsection will be used to derive closed-form performance metrics of the AF relay system to make a comparison with the RIS-aided system.}
	\subsection{Truncation Error}
	To show the effect of infinite series on the performance of the CDF expression of the sum of product of FTR RVs, truncation error is presented in the following.
	By truncating \eqref{CDFFTRSUMPRO} with the first $M$ terms, we have
	{\small \begin{align} \label{CDFTERROR}
	{{\hat F}_Y}(y){=}&\sum\limits_{{j_{1.1}}, \cdots ,{j_{L,1}} = 0}^M   \cdots  \sum\limits_{{j_{1.N}}, \cdots ,{j_{L,N}} = 0}^M  {\prod\limits_{\iota  = 1}^L {\prod\limits_{\ell  = 1}^N {\frac{{{K_{\iota ,\ell }}^{{j_{\iota ,\ell }}}{d_{\iota ,\ell }}_{{j_{\iota ,\ell }}}}}{{{j_{\iota ,\ell }}!}}\frac{{{m_{\iota ,\ell }}^{{m_{\iota ,\ell }}}}}{{\Gamma ({m_{\iota ,\ell }})}}\frac{1}{{\Gamma ({j_{\iota ,\ell }} + 1)}}} } }\notag\\&\times H_{{{1,0:N,1;}} \cdots {{;N,1}}}^{{{0,0:1,N;}} \cdots {{;1,N}}}\left(\!\! {\left. {\begin{array}{*{20}{c}}
	{{y^{ - 1}}\prod\limits_{\ell  = 1}^N {\left( {\sqrt 2 {\sigma _{{{1}},\ell }}} \right)} }\\
	\vdots \\
	{{y^{ - 1}}\prod\limits_{\ell  = 1}^N {\left( {\sqrt 2 {\sigma _{L,\ell }}} \right)} }
	\end{array}} \right|\begin{array}{*{20}{c}}
	{\left( {{{1;1,}} \cdots ,{{1}}} \right):\left\{ {\left( { - {j_{1,n}},0.5} \right)} \right\}_1^N; \cdots ;\left\{ {\left( { - {j_{L,n}},0.5} \right)} \right\}_1^N}\\
	{ - :\left( {0,1} \right); \cdots ;\left( {0,1} \right)}
	\end{array}}\!\! \right).
	\end{align}}
	The truncation error of the area under the ${F_Y}(y)$ with respect to the first	$M$ terms is given by
	\begin{equation}\label{erroe}	
	\varepsilon \left( {{N}} \right) \buildrel \Delta \over = {F_Y}(\infty ) - {{\hat F}_Y}(\infty ).
	\end{equation}
	\begin{lemma}
	By truncating \eqref{CDFTERROR} with $M$ terms, the truncation error of \eqref{erroe} is given as
	\begin{equation}\label{errorobtainRIS}	
	\varepsilon \left( {{N}} \right) \buildrel \Delta \over = 1-\sum\limits_{{j_{1.1}}, \cdots ,{j_{L,1}} = 0}^M  \cdots  \sum\limits_{{j_{1.N}}, \cdots ,{j_{L,N}} = 0}^M  {\prod\limits_{\iota  = 1}^L {\prod\limits_{\ell  = 1}^N {\frac{{{K_{\iota ,\ell }}^{{j_{\iota ,\ell }}}{d_{\iota ,\ell }}_{{j_{\iota ,\ell }}}}}{{{j_{\iota ,\ell }}!}}\frac{{{m_{\iota ,\ell }}^{{m_{\iota ,\ell }}}}}{{\Gamma ({m_{\iota ,\ell }})}}} } } .
	\end{equation}
	It can be shown that $\varepsilon(\infty) \rightarrow 0$ as $M \rightarrow \infty$.
	\end{lemma}
	\begin{IEEEproof}
	The asymptotic expansions of the multivariate Fox's $H$-function in \eqref{CDFTERROR} can be obtained by computing the residue \cite{alhennawi2015closed}. Thus, while $ y \to \infty  $, eq. \eqref{CDFTERROR} can be expressed as
	\begin{align}
	{{\hat F}_Y}(y){=}\sum\limits_{{j_{1.1}}, \cdots ,{j_{L,1}} = 0}^M  \cdots  \sum\limits_{{j_{1.N}}, \cdots ,{j_{L,N}} = 0}^M  {\prod\limits_{\iota  = 1}^L {\prod\limits_{\ell  = 1}^N {\frac{{{K_{\iota ,\ell }}^{{j_{\iota ,\ell }}}{d_{\iota ,\ell }}_{{j_{\iota ,\ell }}}}}{{{j_{\iota ,\ell }}!}}\frac{{{m_{\iota ,\ell }}^{{m_{\iota ,\ell }}}}}{{\Gamma ({m_{\iota ,\ell }})}}} } }.
	\end{align}	
	Furthermore, we have
	\begin{align}
	\sum\limits_{{j_{1.1}}, \cdots ,{j_{L,1}} = 0}^\infty   \cdots  \sum\limits_{{j_{1.N}}, \cdots ,{j_{L,N}} = 0}^\infty  {\prod\limits_{\iota  = 1}^L {\prod\limits_{\ell  = 1}^N {\frac{{{K_{\iota ,\ell }}^{{j_{\iota ,\ell }}}{d_{\iota ,\ell }}_{{j_{\iota ,\ell }}}}}{{{j_{\iota ,\ell }}!}}\frac{{{m_{\iota ,\ell }}^{{m_{\iota ,\ell }}}}}{{\Gamma ({m_{\iota ,\ell }})}}} } }=1.
	\end{align}		
	Then, we complete the proof by using the above result.
	\end{IEEEproof}
	\begin{table}[t]
	\caption{\label{tab1}Minimum Required Terms and Truncation Error for Different Parameters $m_\ell=m$, $K_\ell=K$, $\Delta_\ell=\Delta$, $\sigma^2_\ell=\sigma^2$ and $L$ with $N=2$} .
	\centering
	\begin{tabular}{|c|c|c|}
	\toprule
	\hline
	Parameter & $\varepsilon \left( L \right)$ & ${\color{blue} M}$ \\
	\hline
	$L=1$, $m=5$, $K=3$, $\Delta=0.5$, $\sigma^2=0.5$ & $6.52 \times 10^{-6}$ & 24 \\
	\hline
	$L=2$, $m=5$, $K=3$, $\Delta=0.5$, $\sigma^2=0.5$ & $7.93 \times 10^{-6}$ & 25 \\
	\hline
	$L=2$, $m=25$, $K=3$, $\Delta=0.5$, $\sigma^2=0.5$ & $8.51 \times 10^{-6}$ & 17 \\
	\hline
	$L=3$, $m=5$, $K=3$, $\Delta=0.25$, $\sigma^2=0.5$ & $8.34 \times 10^{-6}$ & 21 \\
	\hline
	\bottomrule
	\end{tabular}
	\end{table}
	
	To demonstrate the convergence of the series in \eqref{errorobtainRIS},
	Table \ref{tab1} depicts the required truncation terms $M$ for different system and channel parameters. With a satisfactory accuracy (e.g., smaller than $10^{-5}$), only less than $30$ terms are needed for all considered cases.
	
	\section{Phase Shift and Power Optimization}\label{optimals}
	{\color{black}We derive the statistics of $\gamma_{{\rm RIS}}$ under the optimal phase shift design of RIS's reflector array, ${\bf \Phi}$, in Section \ref{SFSGA}. In this section, we propose an algorithm to obtain the optimal design of ${\bf \Phi}$. Moreover, to make a fair comparison, we also present a power allocation method for the AF relay system, and derive the statistics of $\gamma_{{\rm AF}}$ under the optimal power allocation scheme.}	
	\subsection{Optimal Phase Shift Design of RIS's Reflector Array}
	To characterize the fundamental performance limit of RIS, we assume that the phase shifts can be continuously varied in $ \left( {0,2\pi } \right] $. {\color{black}Based on \eqref{snrris}, maximizing the SNR ${\gamma _{{\rm RIS}}}$ is equivalent to maximizing ${\left| {\sum\limits_{\ell  = 1}^L {{h_\ell }{g_\ell }{e^{i({\theta _{\ell ,1}} + {\theta _{\ell ,2}} + {\phi _\ell })}}} } \right|^2}$, where ${\phi _\ell }$ is the adjustable phase induced by the $i_{\rm th}$ reflecting element of the RIS. According to \cite[eq. (12)]{basar2019wireless}, it is easy to infer that ${\gamma _{{\rm RIS}}}$ is maximized by eliminating the channel phases (similar to co-phasing in classical maximum ratio combining schemes), i.e., the optimal choice of $\phi _\ell$ that maximizes the instantaneous SNR is ${\theta _{\ell ,1}} + {\theta _{\ell ,2}} + {\phi _\ell }= \phi$ for all $\ell$, where $\phi$ is a constant phase and $\phi \in \left( {{{0}},{{2\pi}}} \right]$.} However, this solution, notably, requires that the channel phases are perfectly known at the RIS. How to perform channel estimation in RIS-aided system is challenging, because the RIS is assumed to be passive, as opposed to, e.g., the AF relay. Moreover, in \cite{han2019large}, the BS is assumed to be equipped with a large uniform linear array; therefore, the problem becomes to measure the AoD and AoA at the RIS. However, the authors in \cite{han2019large} did not obtain a feasible solution to this problem. Herein, we propose a novel and simple algorithm based on the binary search tree \cite{allen1978self}.
	Note that the maximum achievable expectation of the amplitude of the received signal can be expressed as
	\begin{equation}	
	{\mathbb E}\left[ {\left| {\sum\limits_{\ell  = 1}^L {{h_\ell }{g_\ell }{e^{i({\theta _{\ell ,1}} + {\theta _{\ell ,2}} + {\phi _\ell })}}} } \right|} \right]  \le {\mathbb E}\left[ {\sum\limits_{\ell  = 1}^L {{h_\ell }{g_\ell }} } \right]=\sum\limits_{\ell  = 1}^L {\mathbb E}{\left[ {{h_\ell }{g_\ell }} \right]}
	\end{equation}
	where  ${\mathbb E}{\left[ {{h_\ell }{g_\ell }} \right]}$ can be calculated with the aid of \eqref{expdex}.
	
\begin{figure}[t]
	\centering
	\includegraphics[scale=0.8]{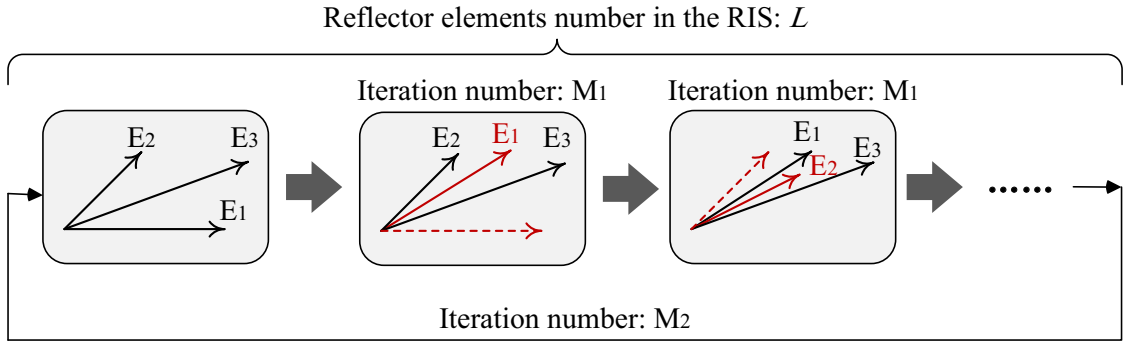}
	\caption{The binary search tree algorithm for RIS-aided system.}
	\label{SuanFa}
	\end{figure}

	Let ${\rm E_{re}} \triangleq {\mathbb E}\left[ {\left| {\sum\limits_{\ell  = 1}^L {{h_\ell }{g_\ell }{e^{i({\theta _{\ell ,1}} + {\theta _{\ell ,2}} + {\phi _\ell })}}} } \right|} \right]$ and ${\rm E_{opt}} \triangleq \sum\limits_{\ell  = 1}^L{\rm E_{\ell}}=\sum\limits_{\ell  = 1}^L {\mathbb E}{\left[ {{h_\ell }{g_\ell }} \right]}$, and we notice that ${\rm E_{re}}$ can be easily measured in the time domain and ${\rm E_{opt}}$ can be calculated with the parameters of FTR fading. Thus, when initially setting up the RIS, we use the binary search tree algorithm to adjust the reflection angle of the elements on the RIS one by one to maximum ${\rm E_{re}}$. For each element, we perform $M_1$ times of search. After searching all the elements, we repeat the same operation $M_2$ times. The entire algorithm flow is shown in Fig. \ref{SuanFa}. {\color{blue}From Fig. \ref{SuanFa}, we can observe that our algorithm can be regarded as $M_1M_2L$ iterations of binary search tree algorithms. Thus, the complexity for each iteration of our algorithm is same as the complexity of binary search tree algorithm that is shown as $\mathcal{O}(log_2(N))$.} This solution does not require to estimate the channel model and does not compromise on the almost passive nature of the RIS. We only need to adjust $\phi_\ell$ and measure the corresponding ${\rm E_{re}}$ in a practical communication system. In simulation experiments, ${\rm E_{opt}}$ can be regarded as an upper bound of ${\rm E_{re}}$ and can be used to test whether our algorithm can converge to the optimal solution.
	
	Based on the above discussion, we present our method to find ${{\bf \Phi} _{{\rm opt}}}$ as Algorithm \ref{1}, where $  {\rm Mod}\left[ {a,b} \right] $ denotes the remainder on division of $a$ by $b$. Thus, with the optimal phase shift design of RIS's reflector array, we can obtain the maximum of ${\gamma _{{\rm RIS}}}$ as \eqref{maxris}.
	\begin{algorithm}[t]
	\caption{The binary search tree algorithm for finding ${{\bf \Phi} _{{\rm opt}}}$.} 
	\label{1}
	\hspace*{0.02in} {\bf Input:}
	input the number of elements on RIS: $L$, iteration numbers: $M_1$ and $M_2$\\
	\hspace*{0.02in} {\bf Output:}
	The phase-shift variable ${\rm \phi_{\ell}}$ $(\ell=1,\cdots,L)$ for each RIS element should be set.
	\begin{algorithmic}[1]
	\State Initialize ${\rm \phi_{LB}}=0$, ${\rm \phi_{UB}}=\pi$ and $\ell_{\rm th}=1$. Set iteration index ${\rm LP_1}={\rm LP_2}={\rm t}=0$.
	\While{${\rm LP_2}< M_2$} 
	\While{$\ell_{\rm th}\le L$}
	\While{${\rm LP_1}< M_1$}
	\State When the phase-shift variable of the $\ell_{\rm th}$ elements is set as ${\rm \phi_{LB}}$ and ${\rm \phi_{UB}}$, we use ${\rm E_{m_1}}$ and ${\rm E_{m_2}}$ to denote the expectation of the amplitude of the received signal measured at the user, respectively.
	\State $ t \leftarrow t + 1 $
	\If{${\rm E_{m_1}}>{\rm E_{m_2}}$}
	\State ${\rm \phi_{LB}} \leftarrow {\rm Mod}\left[{\rm \phi_{LB}}+ {\frac{\pi }{{{2^{t + 1}}}} + 1,2\pi } \right] $ and ${\rm \phi_{UB}} \leftarrow {\rm Mod}\left[{\rm \phi_{LB}} - {\frac{\pi }{{{2^{t}}}} ,2\pi } \right] $.
	\Else \State ${\rm \phi_{UB}} \leftarrow {\rm Mod}\left[{\rm \phi_{UB}}+ {\frac{\pi }{{{2^{t + 1}}}} ,2\pi } \right] $ and ${\rm \phi_{LB}} \leftarrow {\rm Mod}\left[{\rm \phi_{UB}} - {\frac{\pi }{{{2^{t}}}},2\pi } \right] $.	
	\EndIf
	\State ${\rm LP_1}\leftarrow{\rm LP_1+1}$
	\EndWhile
	\State ${\ell_{\rm th}}\leftarrow{\ell_{\rm th}+1}$, $t\leftarrow 0$ and ${\rm LP_1}\leftarrow 0$.		
	\EndWhile
	\State ${\rm LP_2}\leftarrow{\rm LP_2+1}$, ${\rm \phi_{\ell}}\leftarrow {\rm Mod}\left[{\rm \phi_{\ell}}+{\rm \phi_{LB}}\right]$, ${\rm \phi_{LB}}\leftarrow 0$ and ${\rm \phi_{UB}}\leftarrow 0$
	\EndWhile
	\State \Return ${\rm \phi_{\ell}}$ $(\ell=1,\cdots,L)$
	\end{algorithmic}
	\end{algorithm}

{\color{black}
 \begin{lemma}\label{consd}
 	To investigate the convergence of the proposed phase optimization method, we define ${\alpha _{kL + \ell }} \triangleq \theta _{\ell,1} + \theta _{\ell,2}+{\phi _{kL + \ell } }$ $(\ell=1,\cdots,L$ and $k=0,1,\cdots, M_2)$. When $M_1\to \infty$ and $k\to \infty$, we have ${\alpha _{kL + \ell }} \to \phi$ for all $\ell$, where
\begin{equation}\label{afagbv}
\phi  = \frac{{\rm{2}}}{{{L^{\rm{2}}} - L}}\left( {{\alpha _2} + 2{\alpha _3} + 3{a_4} +  \cdots  + \left( {L - 1} \right){a_L}} \right).
\end{equation}
\end{lemma}
\begin{IEEEproof}
See Appendix \ref{AppendixEE}.
\end{IEEEproof}

}
	\begin{figure}[t]
	\begin{minipage}[t]{0.45\linewidth}
	\centering
	\includegraphics[width=1.1\textwidth]{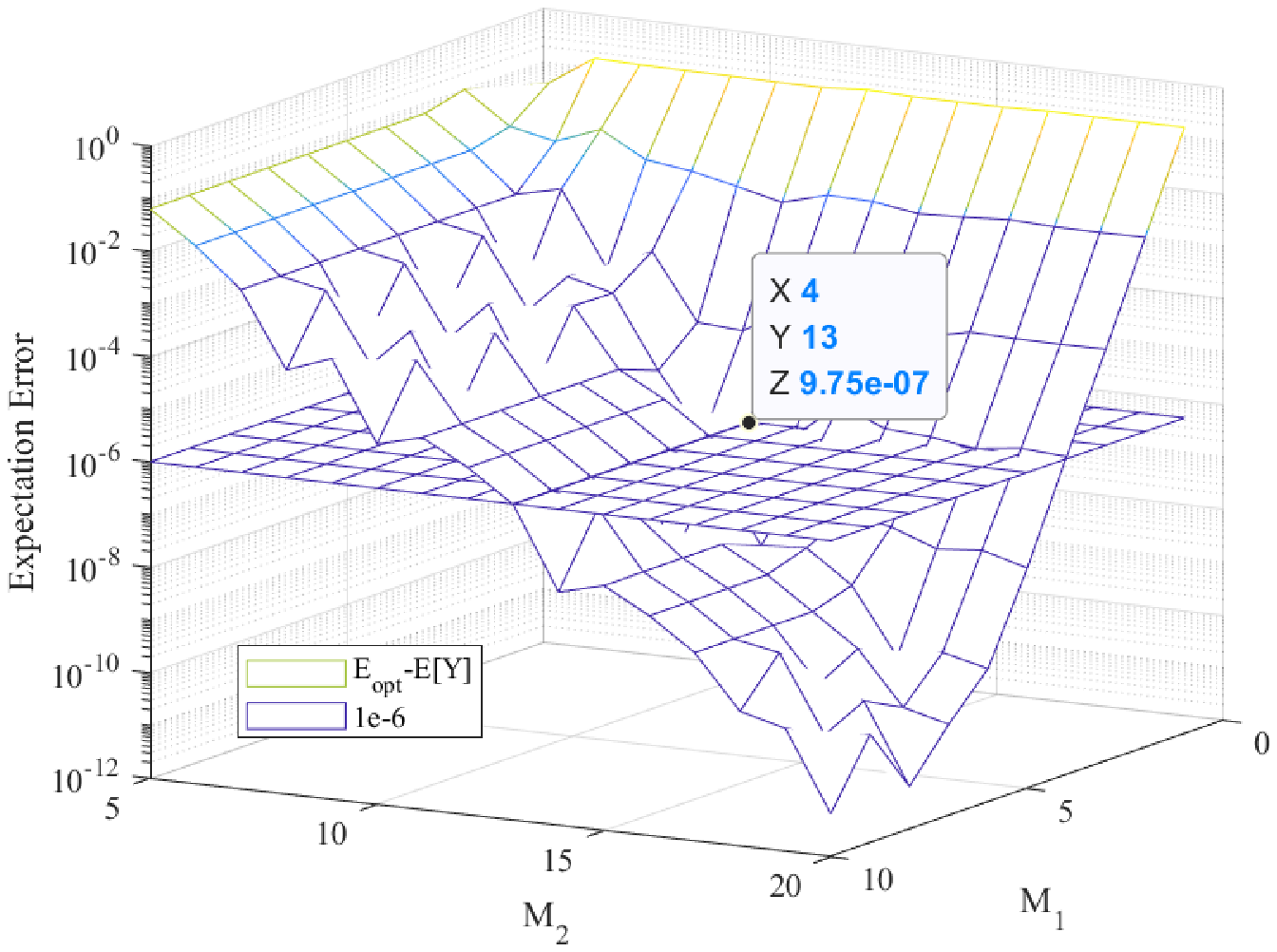}	
	\caption{The expectation error of the proposed phase optimization method versus $M_1$ and $M_2$.}
	\label{EERROR}
	\end{minipage}
	\hfill
	\begin{minipage}[t]{0.45\linewidth}
	\centering
	\includegraphics[width=1.1\textwidth]{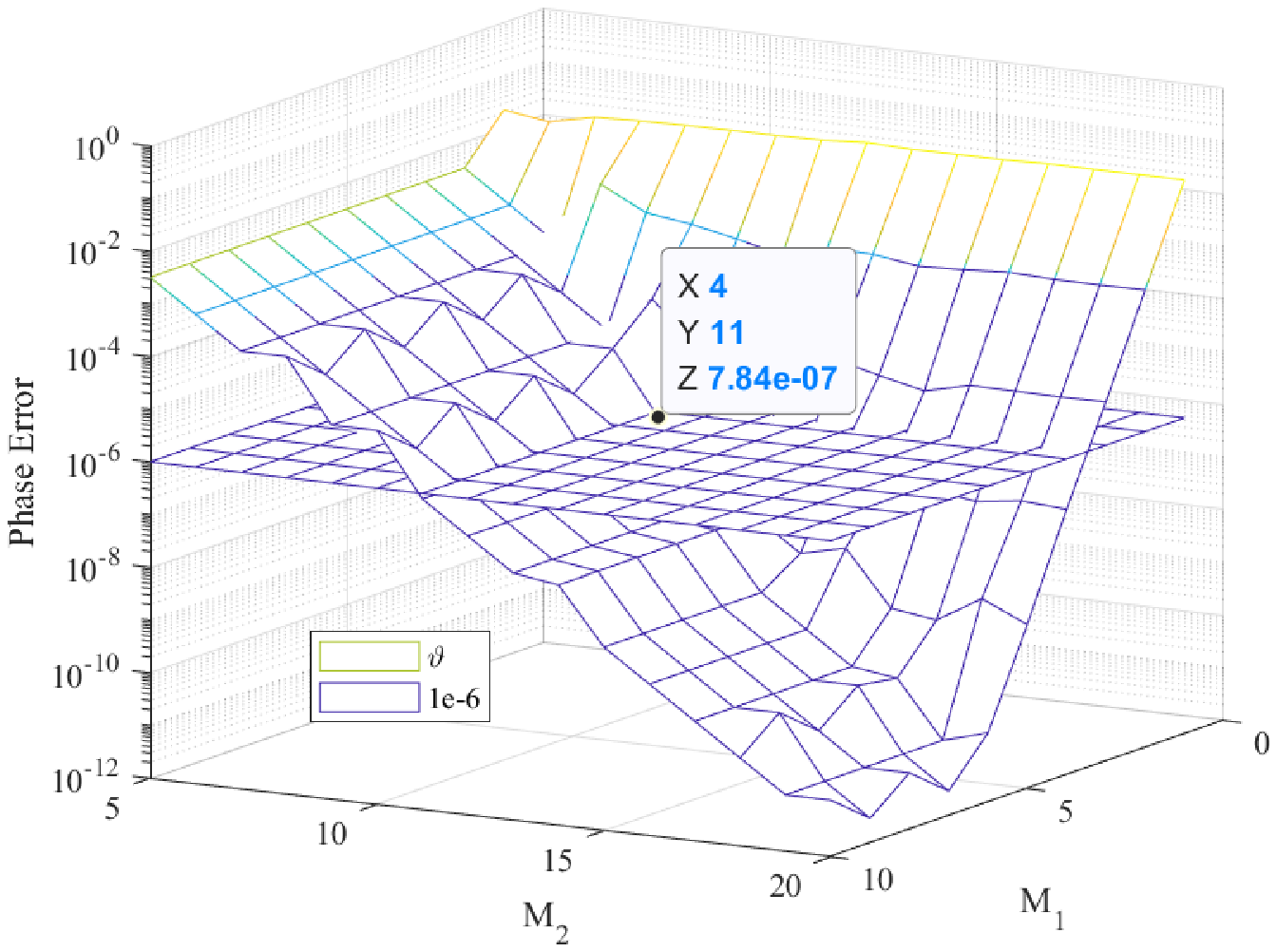}
	\caption{The phase error of the proposed phase optimization method versus $M_1$ and $M_2$.}
	\label{EERROR2}
	\end{minipage}
	\begin{minipage}[t]{0.45\linewidth}
	\centering
	\includegraphics[width=1.1\textwidth]{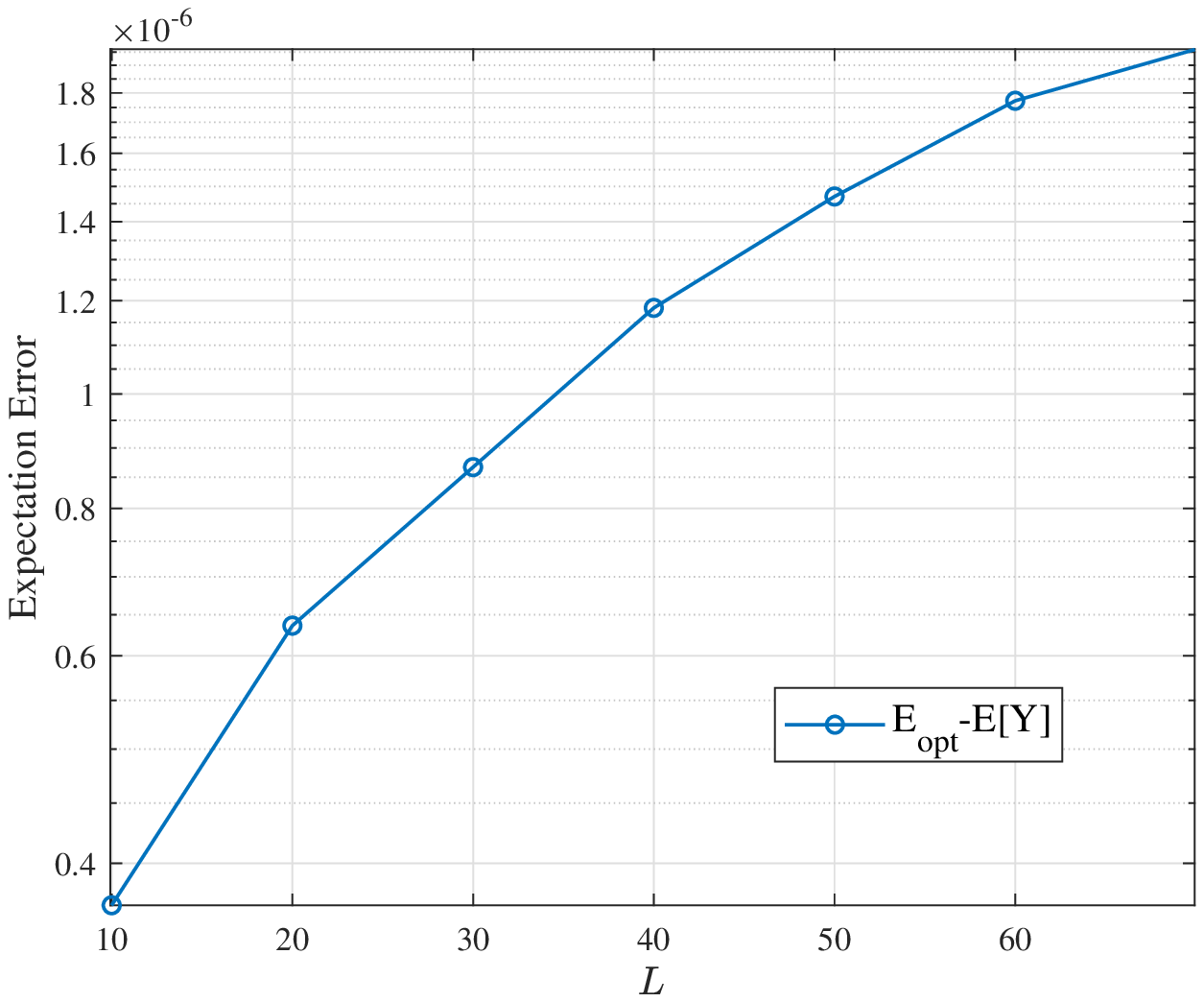}	
	\caption{The expectation error of the proposed phase optimization method versus $L$.}
	\label{EERRORN}
	\end{minipage}
	\hfill
	\begin{minipage}[t]{0.45\linewidth}
	\centering
	\includegraphics[width=1.1\textwidth]{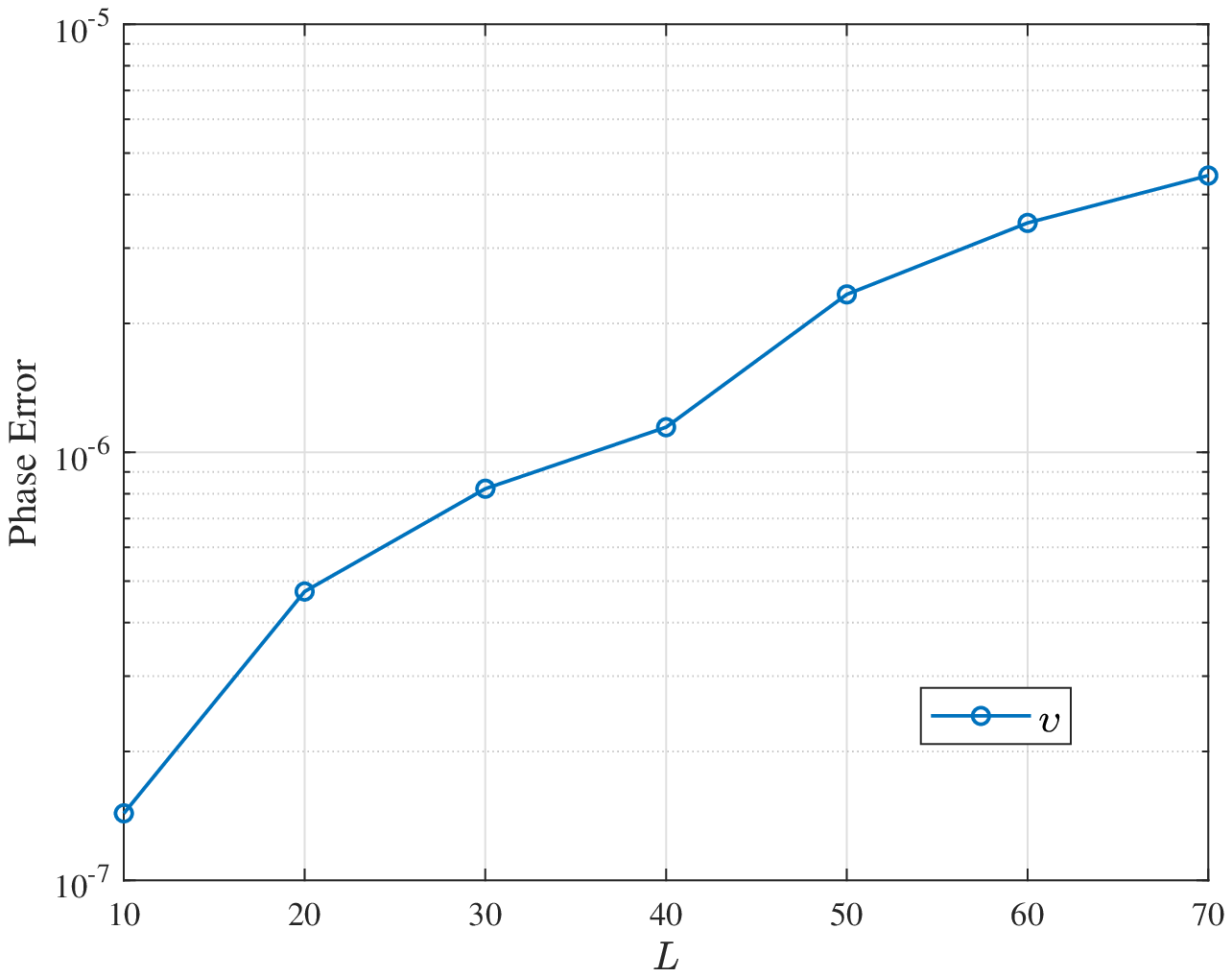}
	\caption{The phase error of the proposed phase optimization method versus $L$.}
	\label{EERRORN2}
	\end{minipage}
	\end{figure}
We define the error of expectation as $\zeta\triangleq{\rm E_{opt}}-{\rm E_{re}}$ and use the variance of ${{\theta _{\ell ,1}} + {\theta _{\ell ,2}} + {\phi _\ell }}$ $(\ell=1,\cdots,L)$ to characterize the phase error. In Figs. \ref{EERROR} and \ref{EERROR2}, we verify the effectiveness of the proposed algorithm for different iteration numbers $M_1$ and $M_2$.  Figures \ref{EERROR} and \ref{EERROR2}, respectively, plot an example of the expectation and phase error, assuming $L=3$, ${\theta _{1 ,1}} + {\theta _{1 ,2}}=\pi/4$, ${\theta _{2,1}} + {\theta _{2,2}}=\pi/2$, ${\theta _{3,1}} + {\theta _{3 ,2}}=\pi/8$, $m_{1,1}=m_{1,2}=10$, $K_{1,1}=K_{1,2}=3$, $\upsilon_{1,1}=\upsilon_{1,2}=10$, $\delta_{1,1}=\delta_{1,2}=0.5$, $m_{2,1}=m_{2,2}=5$, $K_{2,1}=K_{2,2}=5$, $\upsilon_{2,1}=\upsilon_{2,2}=20$, $\delta_{2,1}=\delta_{2,2}=0.5$, $m_{3,1}=m_{3,2}=15$, $K_{3,1}=K_{3,2}=1$, $\upsilon_{3,1}=\upsilon_{3,2}=10$ and $\delta_{3,1}=\delta_{3,2}=0.3$. With the help of Algorithm \ref{1}, we obtain $\phi_1=0.076$, $\phi_2=5.575$, $\phi_3=0.469$ and $\phi=0.86$. Using \eqref{expdex}, we obtain ${\rm E_1}=8.36$, ${\rm E_2}=12.79$, ${\rm E_3}=8.16$ and ${\rm E_{opt}}=29.33$. We can observe that both expectation and phase error decrease fast with the iteration numbers $M_1$ and $M_2$. Furthermore, only less than $50$ iteration numbers are needed for each element to obtain a satisfactory accuracy (e.g., smaller than $10^{-6}$). Moreover, Figures \ref{EERRORN} and \ref{EERRORN2}, respectively, depict the expectation error and phase error of the proposed phase optimization method versus the amount of reflecting elements on the RIS, $L$, with $M_1=4$, $M_2=10$, $m_{\ell,1}=m_{\ell,2}=15$, $K_{\ell,1}=K_{\ell,2}=1$, $\upsilon_{\ell,1}=\upsilon_{\ell,2}=10$ and $\delta_{\ell,1}=\delta_{\ell,2}=0.3$. Considering the searching results may be affected by the initialization, we calculate the average expectation and phase error of $100$ random ${\theta _{\ell,1}}$ and ${\theta _{\ell,2}}$ to obtain Fig. \ref{EERRORN} and \ref{EERRORN2}. As it can be observed, both expectation error and phase error increase with the increase of $L$, but $L$ has only a small influence to the convergence of our phase optimization method. Our algorithm can guarantee to converge to the optimal solution even when the size of the RIS is large. By increasing $M_1$ and $M_2$, more accuracy is achieved upon convergence.
	\subsection{Optimal Power Allocation Scheme for the Amplify-and-Forward Relay System}
	To make a fair comparison, we first select $P_1$ and $P_2$ optimally, while having the same average power $P$ as when using the RIS. Assuming that $P_1$ and $P_2$ are selected under the constraint $P_1+P_2=2P$ and using \eqref{AFDESNR}, we can obtain
	\begin{equation}
	\frac{1}{{{o^{{2}}}{\gamma _{AF}}}}= \left( {\frac{{{1}}}{{{{\left| {{q_2}} \right|}^{{2}}}{P_{{2}}}}} + \frac{{{1}}}{{{{\left| {{q_{{1}}}} \right|}^{{2}}}\left( {{{2}}P - {P_2}} \right)}}} \right).
	\end{equation}
	Thus, we derive the derivative of  ${\frac{1}{{{o ^{{2}}}{\gamma _{AF}}}}} $ as
	\begin{equation}\label{gammadaoshu}
	{\left( {\frac{1}{{{o^{{2}}}{\gamma _{AF}}}}} \right)^\prime } = \frac{{{1}}}{{{{\left| {{q_{{1}}}} \right|}^{{2}}}{{\left( {{{2}}P - {P_2}} \right)}^2}}} - \frac{{{1}}}{{{{\left| {{q_2}} \right|}^{{2}}}{P_{{2}}}^2}}.
	\end{equation}
	With the help of \eqref{gammadaoshu}, we obtain
	\begin{equation}\label{pfenpei}
	{P_1} = \frac{{{2}P\left| {{q_2}} \right|}}{{\left| {{q_{1}}} \right| + \left| {{q_2}} \right|}}, {P_{2}} = \frac{{{2}P\left| {{q_{1}}} \right|}}{{\left| {{q_{1}}} \right| + \left| {{q_2}} \right|}}.
	\end{equation}
	Substituting \eqref{pfenpei} into \eqref{AFDESNR}, we obtain the maximum of $\gamma _{AF}$ as
	\begin{equation}
	\gamma _{AF}^{\max } = \frac{{{{2}}P}}{{{o ^{{2}}}}}\frac{{{{\left| {{q_{{1}}}} \right|}^{{2}}}{{\left| {{q_2}} \right|}^{{2}}}}}{{{{\left( {\left| {{q_{{1}}}} \right|{{ + }}\left| {{q_2}} \right|} \right)}^{{2}}}}}.
	\end{equation}
	{\color{black}Thus, the statistics of the SNR in AF relay system, $\gamma _{AF}$, can be derived as the following corollary.}
	\begin{corr}
	The PDF and CDF expressions of ${\gamma _{AF}^{\max }}$ can be derived as
	{\small \begin{align}
	{f_{\gamma _{AF}^{\max }}}(z) \!=& \frac{{o^2}}{P}\!\sum\limits_{{j_{{1}}} = 0}^\infty  {\!\sum\limits_{{j_{{2}}} = 0}^\infty  {\prod\limits_{\ell  = 1}^2 {\left(\!\! {\frac{{{{m_\ell }^{{m_\ell }}}{K_\ell }^{{j_\ell }}{d_\ell }_{{j_\ell }}}{{{\left( {\frac{{{o^2}}}{{2P}}z} \right)}^{{{1}} + {j_1} + {j_2}}}}}{{\Gamma ({m_\ell }){{j_\ell }!}{{\left( {2{\sigma _\ell }^2} \right)}^{{j_\ell } + 1}}\Gamma ({j_\ell } + 1)}}} \!\!\right)} } }
	H_{0,1:2,0;2,0}^{0,0:0,2;0,2}\left(\!\!\!\! {\left. {\begin{array}{*{20}{c}}
	{\frac{{{{4}}P{\sigma _{{1}}}^2}}{{{o^2}z}}}\\
	{\frac{{{{4}}P{\sigma _{{2}}}^2}}{{{o^2}z}}}
	\end{array}} \!\!\right|\!\!\begin{array}{*{20}{c}}
	{ - :\left( {{{2}},{{2}}} \right)\left( {3 + 2{j_1},{{2}}} \right);\left( {{{2}},{{2}}} \right)\left( {3 + 2{j_2},{{2}}} \right)}\\
	{\left( {4 + 2{j_1} + 2{j_2};{{2}},{{2}}} \right): - }
	\end{array}}\!\!\! \right),
	\end{align}}
	\vspace{-5mm}
	{\small \begin{align}
	{F_{\gamma _{AF}^{\max }}}\!\!\left( z \right) =& 1 \!- \! \frac{{{m_2}^{{m_2}}}}{{\Gamma ({m_2})}}\!\!\sum\limits_{{j_2} = 0}^\infty  {\frac{{{K_2}^{{j_2}}{d_2}_{{j_2}}}}{{{j_2}!}}} \frac{1}{{\Gamma ({j_2} + 1)}}\Gamma \left(\!\! {1 \!+\! {j_2},\frac{{{o^2}z}}{{{{4}}P{\sigma _{{2}}}^2}}} \!\!\right)
	\!+\! 2\sum\limits_{{j_1} = 0}^\infty  {\sum\limits_{{j_2} = 0}^\infty  {\prod\limits_{\ell  = 1}^2 {\left( {\frac{{{m_\ell }^{{m_\ell }}}}{{\Gamma ({m_\ell })}}\frac{{{K_\ell }^{{j_\ell }}{d_\ell }_{{j_\ell }}}}{{{j_\ell }!}}\frac{{{1}}}{{\Gamma ({j_\ell } + 1)}}} \right)} } } \frac{{{{\left( {{o^2}z} \right)}^{1 + {j_2}}}}}{{{{\left( {4P{\sigma _2}^2} \right)}^{{j_2} + 1}}}}\notag\\&
	\times H_{{{0}},{{1:3,1;2,0}}}^{{{0}},{{0:1,2;0,2}}}\left( {\left. {\begin{array}{*{20}{c}}
	{\frac{{{{4}}P{\sigma _{{1}}}^2}}{{{o^2}z}}}\\
	{\frac{{{{4}}P{\sigma _{{2}}}^2}}{{{o^2}z}}}
	\end{array}} \right|\begin{array}{*{20}{c}}
	{ - :\left( {0,2} \right)\left( { - {j_1},1} \right)\left( {1,1} \right);\left( {1,1} \right)\left( {3 + 2{j_2},2} \right)}\\
	{\left( {1 + 2{j_2};2,2} \right):\left( {0,1} \right); - }
	\end{array}} \right).
	\end{align}	}
	\end{corr}
	\begin{IEEEproof}Substituting $R = \sqrt \gamma $ into \eqref{PDFFTR} and \eqref{CDFFTR}, the PDF and CDF of $\frac{{\left| {{q_{{1}}}} \right|\left| {{q_2}} \right|}}{{\left| {{q_{{1}}}} \right|{{ + }}\left| {{q_2}} \right|}}$ can be derived easily following the similar procedures as in Theorem \ref{them4}. Employing a transformation of RVs, the CDF of $\gamma _{AF}^{\max }$ can be derived to complete the proof.
	\end{IEEEproof}
	
	\section{Performance Analysis of Two Systems}\label{perfos}
	\subsection{Outage Probability}
	The outage probability $P$ is defined as the probability that the received SNR per signal falls below a given threshold $\gamma_{th}$. Thus, the OP can be obtained as ${P_{{\rm{out}}}} = P\left( {\gamma < {\gamma _{{\rm{th}}}}} \right){\rm{ = }}F_{\gamma}\left( {{\gamma _{{\rm{th}}}}} \right)$. {\color{black}Therefore, the OP of the RIS-aided and AF relay system can be directly evaluated by using \eqref{cor1cdf} and \eqref{CDFAFOUTH}, respectively. From \eqref{cor1cdf} and \eqref{CDFAFOUTH}, as expected, we can observe the outage probability decreases when the multipath parameters $m$ increase due to the improved communication conditions for both RIS-aided system and AF relay system.}
	\subsection{Numerical Results of Outage Probability}
	We first present the numerical results for the system average outage probability and the number of Monte Carlo iterations is $10^6$.
	\begin{figure}[t]
	\begin{minipage}[t]{0.45\linewidth}
	\centering
	\includegraphics[width=1.1\textwidth]{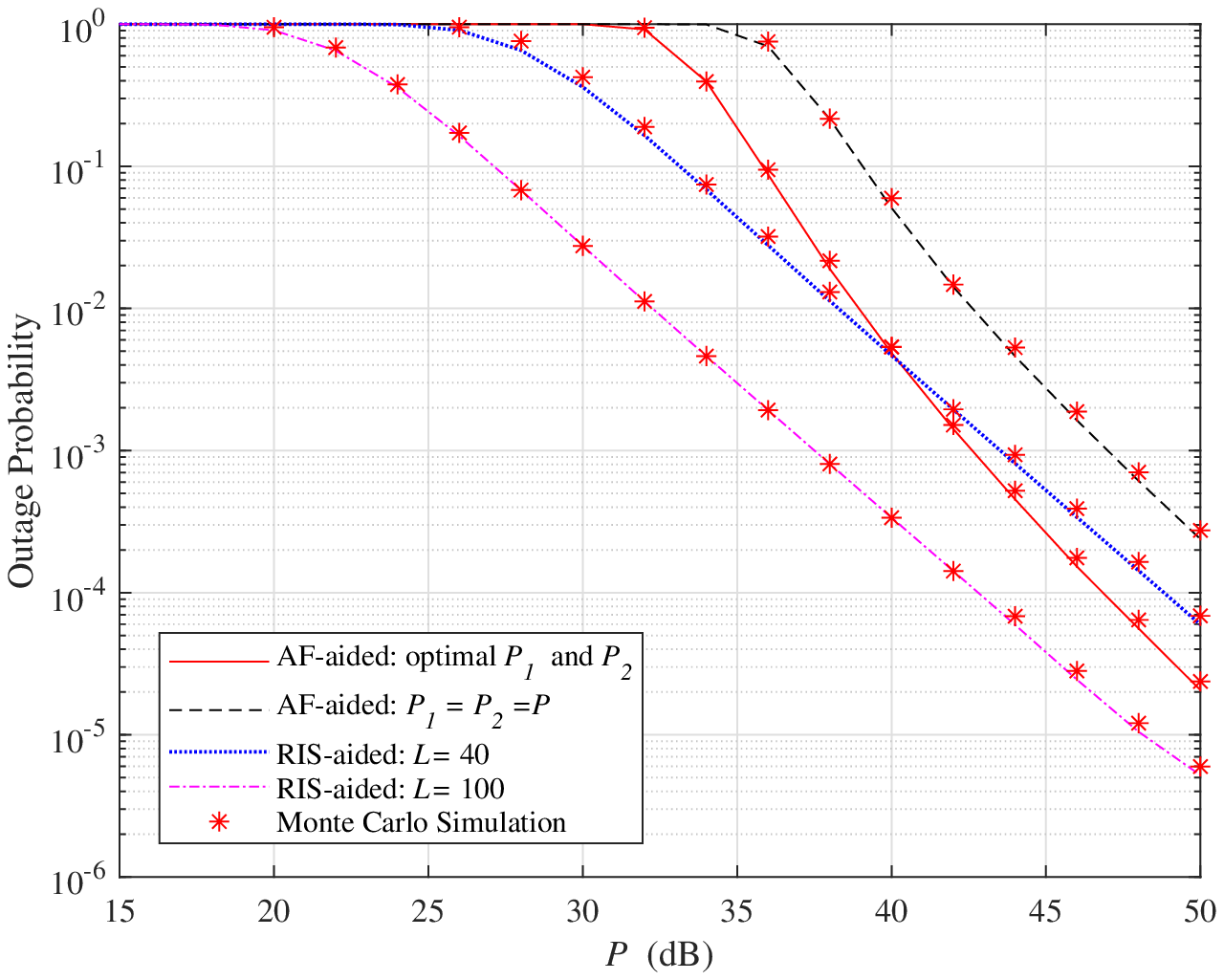}	
	\caption{Outage probability of the RIS-aided and AF relay systems versus the transmit power with $\sigma_N=0$ ${\rm dB}$, $r_{th}=1$, $m_{\ell,1}=5$, $m_{\ell,2}=10$, $K_{\ell,1}=5$, $K_{\ell,2}=7$, $\delta_{\ell,1}=0.5$, $\delta_{\ell,1}=0.7$, $ \upsilon=-60$ ${\rm dB}$.}
	\label{OP1}
	\end{minipage}
	\hfill
	\begin{minipage}[t]{0.45\linewidth}
	\centering
	\includegraphics[width=1.1\textwidth]{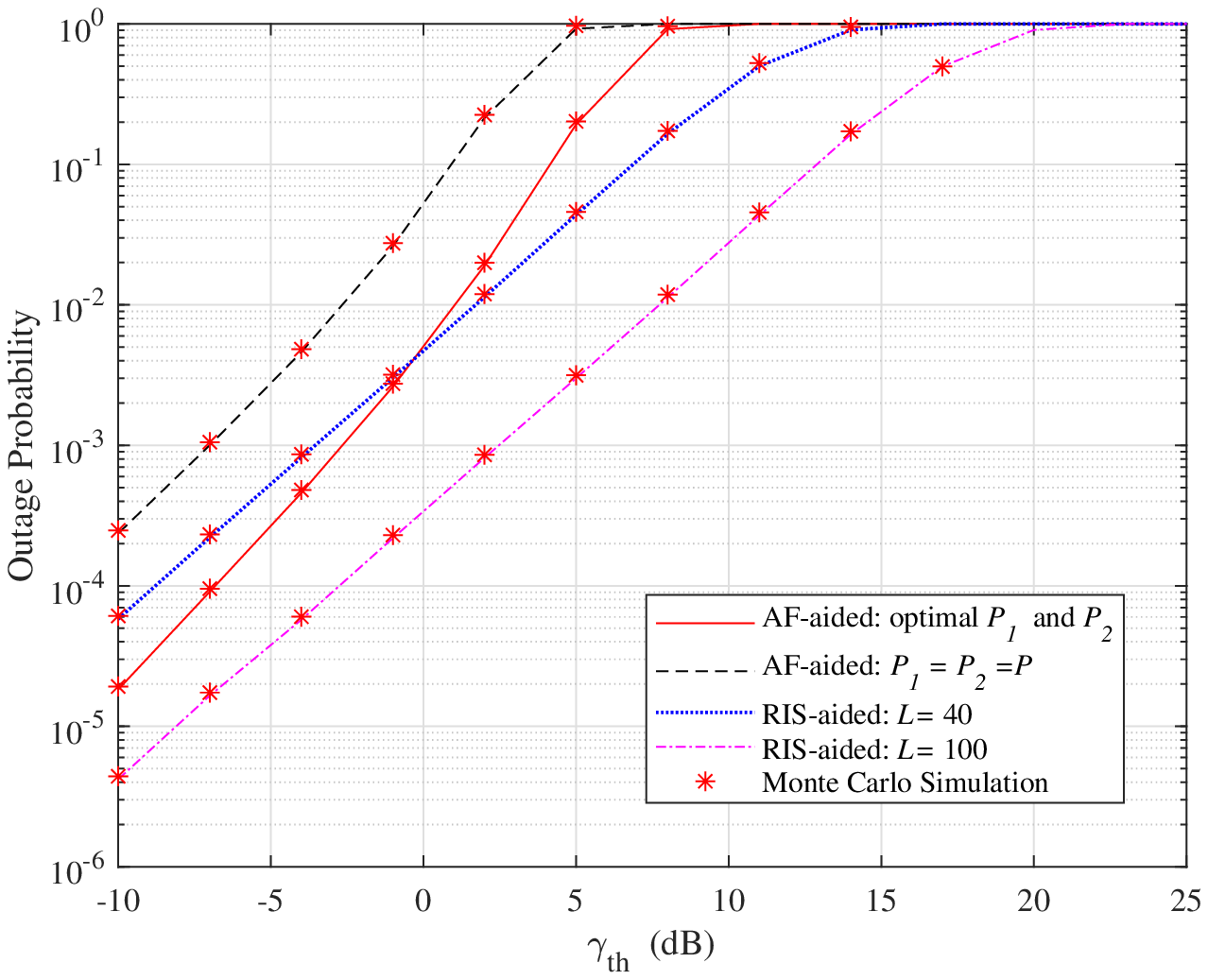}
	\caption{Outage probability of the RIS-aided and AF relay systems versus the threshold SNR with $\sigma_N=0$, $P_{um}=20$ ${\rm dB}$, $m_{\ell,1}=5$, $m_{\ell,2}=10$, $K_{\ell,1}=5$, $K_{\ell,2}=7$, $\delta_{\ell,1}=0.5$, $\delta_{\ell,1}=0.7$, $ \upsilon=-20$ ${\rm dB}$..}
	\label{OP2}
	\end{minipage}
	\end{figure}
	
	Figure \ref{OP1} depicts the OP performance of the RIS-aided and the AF relay system versus the transmit power with $\sigma_N=0$, ${\rm dB}$, $r_{th}=1$, $m_{\ell,1}=5$, $m_{\ell,2}=10$, $K_{\ell,1}=5$, $K_{\ell,2}=7$, $\delta_{\ell,1}=0.5$, $\delta_{\ell,1}=0.7$, $ \upsilon=-60$ ${\rm dB}$. As it can be observed, for the RIS-aided system, the OP decreases as the transmit power and $L$ increase. In addition, using the optimal power allocation scheme, we can see that the AF relay system has lower OP than the RIS-aided system with $40$ reflecting elements and $P_{um}>40$ ${\rm dB}$, and decays faster than RIS-aided system. If the RIS is equipped with $100$ elements, the OP of RIS-aided system is always lower then the AF relay system. Considering that the RIS is usually equipped with hundreds of elements, it is of practical interest to employ RIS in mmWave communications. Furthermore, it is easy to observe from the figure that a good agreement exists between the analytical and simulation results, which justifies the correctness of proposed analytical expressions.
	
	Figure \ref{OP2} illustrates the OP of the two considered system versus the threshold $\gamma_{th}$ with $\sigma_N=0$, $P_{um}=20$ ${\rm dB}$, $m_{\ell,1}=5$, $m_{\ell,2}=10$, $K_{\ell,1}=5$, $K_{\ell,2}=7$, $\delta_{\ell,1}=0.5$, $\delta_{\ell,1}=0.7$, $ \upsilon=-20$ ${\rm dB}$. An interesting insight is that the OP of AF relay system increases faster than that of the RIS-aided system. This implies that the RIS can better adapt to the wireless requirements than the AF relay. Again, under the set of channel parameters, the RIS equipped with $100$ reflecting elements can achieve a lower OP than the AF relay. The reason is that RIS has superior capability in manipulating electromagnetic waves, thus it can provide space-intensive and reliable communication, by enabling the wireless channel to exhibit a line-of-sight. {\color{blue}Moreover, we can observe that the the RIS equipped with 40 reflecting elements achieve a higher OP then the AF relay in the low threshold SNR region. This is because the relay can actively process the received signal and re-transmit an amplified signal, while the RIS only passively reflects the signal without amplification.}
	\subsection{Average Bit-Error Probability}
	Another important performance metric frequently applied is the ABEP, which is given by \cite{zhang2017new}
	\begin{equation}\label{BERDINGYI}	
	{P_e} = \frac{{{q^p}}}{{2\Gamma (p)}}\int_0^\infty  {{z^{p - 1}}} {e^{ - qz}}{F_Z}(z)dz
	\end{equation}
	where $p$ and $q$ denote the modulation-specific parameters for binary modulation schemes, respectively. For instance, $(p, q) = (0.5, 1)$ denotes the binary shift keying (BPSK), $(p, q) = (0.5, 0.5)$ for coherent binary frequency shift keying, and $(p, q) = (1, 1)$ for differential BPSK.
	\begin{prop}\label{sfagea}
	The exact ABEP of RIS-aided system can be expressed	as
	\begin{align}	\label{BERRISZUIZHONG}
	{P_e} =&\frac{{{1}}}{{2\Gamma (p)}}\sum\limits_{{j_{1.1}}, \cdots ,{j_{L,1}} = 0}^\infty  \sum\limits_{{j_{1,2}}, \cdots ,{j_{L,2}} = 0}^\infty  {\prod\limits_{\iota  = 1}^L {\prod\limits_{\ell  = 1}^2 {\frac{{{K_{\iota ,\ell }}^{{j_{\iota ,\ell }}}{d_{\iota ,\ell }}_{{j_{\iota ,\ell }}}}}{{{j_{\iota ,\ell }}!}}\frac{{{m_{\iota ,\ell }}^{{m_{\iota ,\ell }}}}}{{\Gamma ({m_{\iota ,\ell }})}}\frac{1}{{\Gamma ({j_{\iota ,\ell }} + 1)}}} } } \notag\\
	\times& H_{2,0:1,1; \cdots ;1,1}^{0,1:1,1; \cdots ;1,1}\left( {\left. {\begin{array}{*{20}{c}}
	{{{\left( {\frac{qP}{{{o^2}}}} \right)}^{\frac{1}{2}}}\prod\limits_{\ell  = 1}^2 {\left( {\sqrt 2 {\sigma _{1,\ell }}} \right)} }\\
	\vdots \\
	{{{\left( {\frac{qP}{{{o^2}}}} \right)}^{\frac{1}{2}}}\prod\limits_{\ell  = 1}^2 {\left( {\sqrt 2 {\sigma _{L,\ell }}} \right)} }
	\end{array}} \right|\begin{array}{*{20}{c}}
	{\left( {1 - p; - 0.5, \cdots , - 0.5} \right)\left( {1;1, \cdots ,1} \right):\eta }\\
	{ - :\left( {0,1} \right); \cdots ;\left( {0,1} \right)}
	\end{array}} \right),
	\end{align}
	where $\eta  = \left\{ {\left( { - {j_{1,n}},0.5} \right)} \right\}_1^2; \cdots ;\left\{ {\left( { - {j_{L,n}},0.5} \right)} \right\}_1^2.$
	\end{prop}
	\begin{IEEEproof}
	See Appendix \ref{AppendixE}.
	\end{IEEEproof}
	{\color{black}From \eqref{BERRISZUIZHONG}, we can observe that ABEP decreases when the multipath parameters $m$ is improved. Furthermore, a RIS equipped with more reflecting elements will also make the ABEP lower.}
	{\color{black}To make a comparison, for the considered AF relay system, the ABEP with any $P_1$ and $P_2$ or with optimally selected $P_1$ and $P_2$ are both derived in the flowing propositions.}
	\begin{prop}\label{sfdsff}
	The exact ABEP of AF relay system with any $P_1$ and $P_2$ can be expressed as
	{\small \begin{align}\label{BERAF}
	P_e^{AF} =& \frac{1}{2} -\frac{{{m_{{2}}}^{{m_{{2}}}}}}{{2\Gamma (p)\Gamma ({m_{{2}}})}}\sum\limits_{{j_{{2}}} = 0}^\infty  {\frac{{{K_{{2}}}^{{j_{{2}}}}{d_{{2}}}_{{j_{{2}}}}\Gamma \left( {p + 1 + {j_2}} \right)}}{{{j_{{2}}}!\Gamma ({j_{{2}}} + 1)}}} \frac{{{}_2{F_1}\left( {1,p + 1 + {j_2};p + 1;\frac{{2{\sigma _2}^2{P_2}q}}{{{o^2} + 2{\sigma _2}^2{P_2}q}}} \right)}}{{{{\left( {\frac{{{o^2}}}{{2{P_2}{\sigma _2}^2q}} + 1} \right)}^{ - p}}p{{\left( {{{1}} + \frac{{2{P_2}{\sigma _2}^2q}}{{{o^2}}}} \right)}^{1 + {j_2}}}}}
	\notag\\&
	+ \frac{{{q^{ - 1 - {j_2}}}}}{{2\Gamma (p)}}\sum\limits_{{j_{{1}}} = 0}^\infty  {\sum\limits_{{j_{{2}}} = 0}^\infty  {\prod\limits_{\ell  = 1}^2 {\left( {\frac{{{m_\ell }^{{m_\ell }}}}{{\Gamma ({m_\ell })}}\frac{{{K_\ell }^{{j_\ell }}{d_\ell }_{{j_\ell }}}}{{{j_\ell }!}}\frac{{{1}}}{{\Gamma ({j_\ell } + 1)}}} \right)} } } {\left( {\frac{{{o^2}}}{{2{\sigma _{{2}}}^2{P_{{2}}}}}} \right)^{{j_{{2}}} + 1}}\notag\\&
	\times H_{1,1:{{3,1;2,0}}}^{0,1:{{1,2;0,2}}}\left( {\left. {\begin{array}{*{20}{c}}
	{\frac{{2{\sigma _1}^2{P_1}q}}{{{o^2}}}}\\
	{\frac{{2{\sigma _2}^2{P_2}q}}{{{o^2}}}}
	\end{array}} \right|\begin{array}{*{20}{c}}
	{\left( { - {j_2} - p; - 1, - 1} \right):\left( { - {j_1},1} \right)\left( {0,1} \right)\left( {0,1} \right);\left( {2 + {j_2},1} \right)\left( {1,1} \right)}\\
	{\left( {1 + {j_2};1,1} \right):\left( {1,1} \right); - }
	\end{array}} \right).
	\end{align}}
	\end{prop}
	\begin{IEEEproof}
	See Appendix \ref{AppendixF}.
	\end{IEEEproof}
	\begin{prop}
	The exact ABEP of AF relay system with optimally selected $P_1$ and $P_2$ can be derived as
	{\small \begin{align}\label{BERAFmax}
	{P_e^{AF'}} =& \frac{1}{2} -\frac{{{m_{{2}}}^{{m_{{2}}}}}}{{2\Gamma (p)\Gamma ({m_{{2}}})}}\sum\limits_{{j_{{2}}} = 0}^\infty  {\frac{{{K_{{2}}}^{{j_{{2}}}}{d_{{2}}}_{{j_{{2}}}}\Gamma \left( {p + 1 + {j_2}} \right)}}{{{j_{{2}}}!\Gamma ({j_{{2}}} + 1)}}} \frac{{{}_2{F_1}\left( {1,p + 1 + {j_2};p + 1;\frac{{4{\sigma _2}^2{P}q}}{{{o^2} + 4{\sigma _2}^2{P}q}}} \right)}}{{{{\left( {\frac{{{o^2}}}{{{4P}{\sigma _2}^2q}} + 1} \right)}^{ - p}}p{{\left( {{{1}} + \frac{{4P{\sigma _2}^2q}}{{{o^2}}}} \right)}^{1 + {j_2}}}}}
	\notag\\&
	+ \frac{{{q^{ - 1 - {j_2}}}}}{{\Gamma (p)}}\sum\limits_{{j_{{1}}} = 0}^\infty  {\sum\limits_{{j_{{2}}} = 0}^\infty  {\prod\limits_{\ell  = 1}^2 {\left( {\frac{{{m_\ell }^{{m_\ell }}}}{{\Gamma ({m_\ell })}}\frac{{{K_\ell }^{{j_\ell }}{d_\ell }_{{j_\ell }}}}{{{j_\ell }!}}\frac{{{1}}}{{\Gamma ({j_\ell } + 1)}}} \right)} } } {\left( {\frac{{{o^2}}}{{4P{\sigma _{{2}}}^2{}}}} \right)^{{j_{{2}}} + 1}}\notag\\&
	\times H_{1,1:{{3,1;2,0}}}^{0,1:{{1,2;0,2}}}\left( {\left. {\begin{array}{*{20}{c}}
	{\frac{{4{\sigma _1}^2{P}q}}{{{o^2}}}}\\
	{\frac{{4{\sigma _2}^2{P}q}}{{{o^2}}}}
	\end{array}} \right|\begin{array}{*{20}{c}}
	{\left( { - {j_2} - p; - 1, - 1} \right):\left( { - {j_1},1} \right)\left( {0,1} \right)\left( {0,1} \right);\left( {2 + {j_2},1} \right)\left( {1,1} \right)}\\
	{\left( {1 + {j_2};1,1} \right):\left( {1,1} \right); - }
	\end{array}} \right).
	\end{align}}
	\end{prop}
	\begin{IEEEproof}
	Following similar steps as in Proposition \ref{sfdsff}, we obtain \eqref{BERAFmax} to complete the proof.
	\end{IEEEproof}
	{\color{black}From \eqref{BERAF} and \eqref{BERAFmax}, as expected, we can observe that the ABEP of AF relay system with optimally selected $P_1$ and $P_2$ can be regarded as a low bound of the ABEP of AF relay system with any $P_1$ and $P_2$. Furthermore, the multipath parameters $m$ have a positive effect.}
	\subsection{Numerical Results of Average Bit-Error Probability}
	\begin{figure}[t]
	\begin{minipage}[t]{0.45\linewidth}
	\centering
	\includegraphics[width=1.1\textwidth]{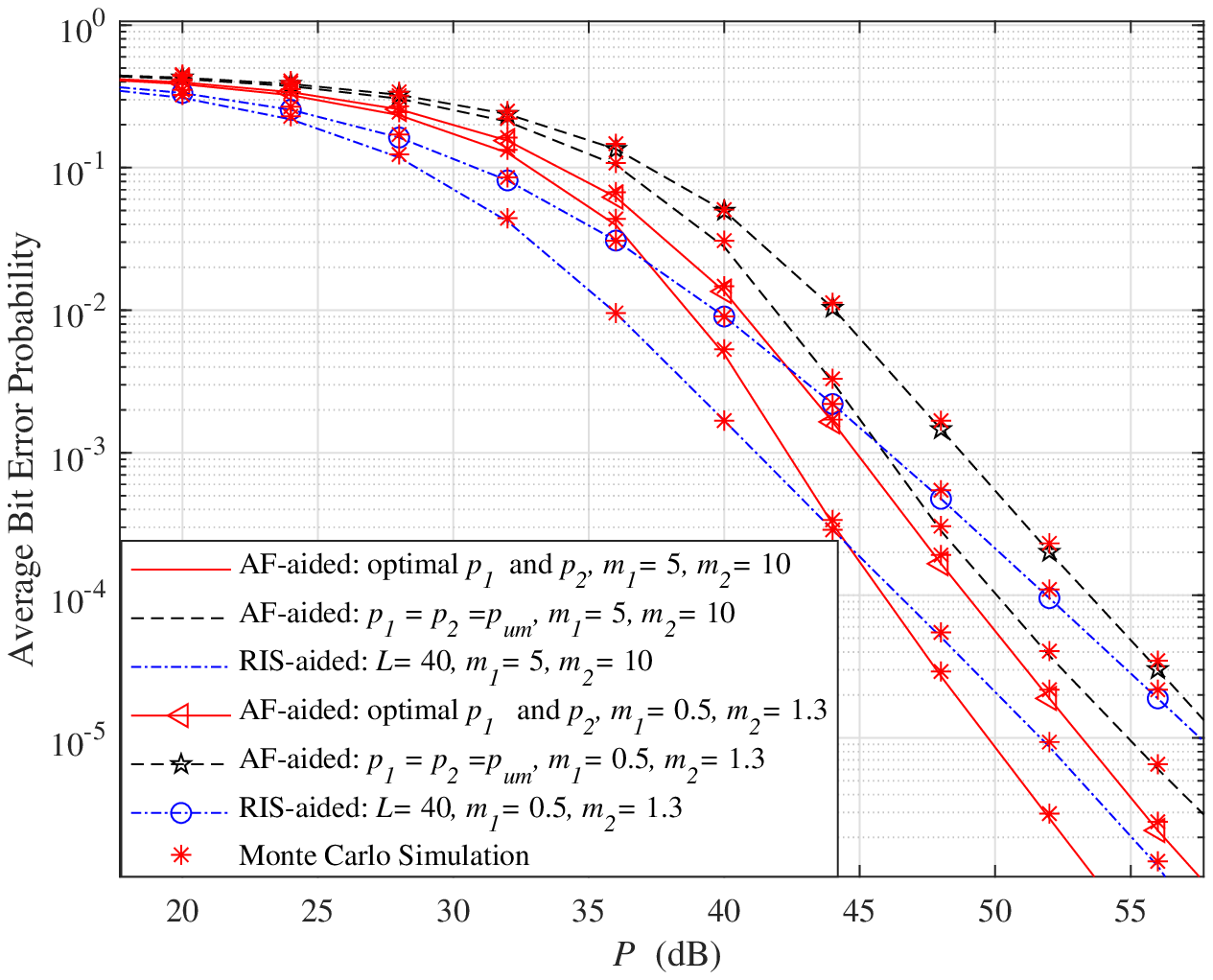}	
	\caption{Average Bit Error Probability of the RIS-aided and AF relay systems versus the transmit power with $\sigma_N=0$, $ \upsilon=-60$ ${\rm dB}$, $K_{\ell,1}=5$, $K_{\ell,2}=7$, $\delta_{\ell,1}=0.5$, $\delta_{\ell,1}=0.7$ and different $m_{\ell,1}$ and $m_{\ell,2}$.}
	\label{ABEP1}
	\end{minipage}
	\hfill
	\begin{minipage}[t]{0.45\linewidth}
	\centering
	\includegraphics[width=1.1\textwidth]{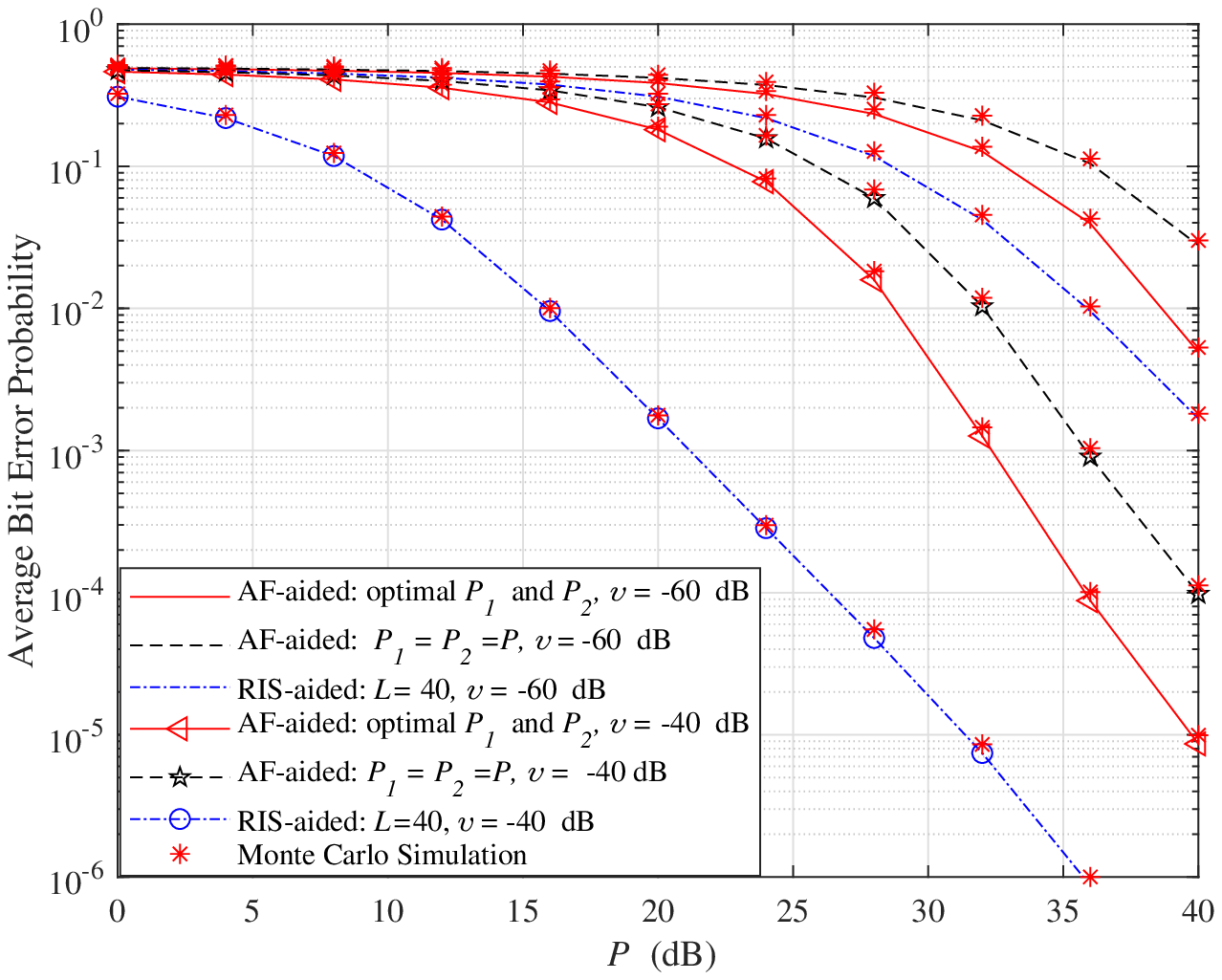}
	\caption{Average Bit Error Probability of the RIS-aided and AF relay systems versus the transmit power with $\sigma_N=0$, $m_{\ell,1}=5$, $m_{\ell,2}=10$, $K_{\ell,1}=5$, $K_{\ell,2}=7$, $\delta_{\ell,1}=0.5$ and $\delta_{\ell,1}=0.7$.}
	\label{ABEP2}
	\end{minipage}
	\end{figure}
		
	Figure \ref{ABEP1} plots the ABEP of the RIS-aided and AF relay systems versus the transmit power with $\sigma_N=0$, $ \upsilon=-60$ ${\rm dB}$, $K_{\ell,1}=5$, $K_{\ell,2}=7$, $\delta_{\ell,1}=0.5$, $\delta_{\ell,1}=0.7$ and different $m_{\ell,1}$ and $m_{\ell,2}$. Obviously, the ABEP decreases as the increase of transmit power for both systems. Similar to the OP, the ABEP of AF relay system decreases faster than the one of the RIS-aided system. Moreover, the curves show the impact of different values of shaping parameters $m_{\ell,1}$ and $m_{\ell,2}$. As $m_{\ell,2}$ is decreased, the ABEP  decreases because of the worse channel conditions. {\color{blue}Furthermore, we can observe that the AF relay system can outperform the RIS-aided system with the same channel conditions in the high transmit power region. This is because the optimal power allocation scheme is used in the AF relay system, and the RIS just reflect signals passively. Thus, because of the low channel gain, RIS need more elements to be competitive in the high transmit power region. Moreover, the figure shows that the RIS equipped with $40$ reflecting elements can always achieve a lower ABEP than the AF relay without the power allocation scheme.}
	
	As shown in Fig. \ref{ABEP2}, with $\sigma_N=0$, $m_{\ell,1}=5$, $m_{\ell,2}=10$, $K_{\ell,1}=5$, $K_{\ell,2}=7$, $\delta_{\ell,1}=0.5$ and $\delta_{\ell,1}=0.7$, the received SNR undergoing the FTR fading channel, $ \upsilon$, has a great impact on the performance of the RIS-aided system. We can observe that the ABEP of the RIS-aided system decreases faster then the AF relay system. It is important to conclude that, as the channel conditions improve, the performance of the RIS-aided system can outperform the AF relay system with the same transmit power. The RIS equipped with only $40$ elements can also provide a lower ABEP than the AF relay. This is due to the reason that the RIS can effectively change the phase of reflected signals without buffering or processing the incoming signals, and the received signal can be enhanced by adjusting the phase shift of each element on the RIS.
	\section{Conclusions}\label{cons}
	In this paper, we made a comprehensive comparison between the RIS-aided and AF relay systems over realistic mmWave channels. We first derived new statistical characterizations of the product of independent FTR RVs and the sum of product of FTR RVs, and used the results to obtain the exact PDF and CDF expressions of the RIS-aided system. In addition, the characterizations of end-to-end SNR of the AF relay system assuming either ideal or non-ideal hardware were obtained. We proposed a novel and simple way to obtain the optimal phases at the RIS, and we also proposed the optimal power allocation scheme for AF relay system. Then, the OP and the ABEP of the two systems were presented to study their performances. In particular, our results show that when the transmit power is low or the threshold is high, the RIS-aided system can achieve the same performance as the AF relay system using a small number reflecting elements. Consequently, compared with the AF relay, RIS is more suitable for mmWave communications.
	\begin{appendices}
	\section{Proof of Theorem \ref{APPAT}}\label{AppendixA} 
	\renewcommand{\theequation}{A-\arabic{equation}}
	\setcounter{equation}{0}
	\subsubsection{Proof of PDF}
	The PDF of $X$ can be formulated as 
	{\small\begin{equation}\label{PDFchushiA}
	 {f_X}(x) = \frac{1}{x}\frac{1}{{2\pi i}}\int_{\cal L} {{\mathbb E}\left[ {{X^s}} \right]{x^{ - s}}} {\rm d}s
	\end{equation}}
	where the integration path of  ${\cal L}$ goes from $\sigma -\infty j$ to $\sigma+\infty j$ and $\sigma  \in \mathbb{R}$.
	By inserting \eqref{moment} into \eqref{PDFchushiA} and carrying
	out some algebraic manipulations, we can re-write \eqref{PDFchushiA}
	{\small\begin{equation}
	 {f_X}(x) = \frac{1}{x}\frac{1}{{2\pi i}}\prod\limits_{\ell  = 1}^N {\frac{{{m_\ell }^{{m_\ell }}}}{{\Gamma ({m_\ell })}}\sum\limits_{{j_\ell } = 0}^\infty  {\frac{{{K_\ell }^{{j_\ell }}{d_\ell }_{{j_\ell }}}}{{{j_\ell }!}}} \frac{2}{{\Gamma ({j_\ell } + 1)}}} \int_{\cal L} {\prod\limits_{\ell  = 1}^N {\Gamma \left( {1 + {j_\ell } + t} \right)} {{\left( {{x^2}\prod\limits_{\ell  = 1}^N {\left( {\frac{1}{{2{\sigma _\ell }^2}}} \right)} } \right)}^{ - t}}} {\rm d}t.
	\end{equation}}
	With the help of \cite[eq. (9.301)]{gradshteyn2007}, we can obtain \eqref{PDFFINAL1}.
	\subsubsection{Proof of CDF}
	The CDF of $X$ can be expressed as
	{\small\begin{equation}\label{CDFCHUSHI1}
	 {F_X}(x) = \int_0^x {{f_r}} (r){\rm d}r.
	\end{equation}}
	Substituting \eqref{PDFFINAL1} into \eqref{CDFCHUSHI1}, we obtain
	{\small\begin{equation}\label{CDFGUOCHENG1}
	 {F_X}\left( x \right) = \prod\limits_{\ell  = 1}^N {\frac{{{m_\ell }^{{m_\ell }}}}{{\Gamma ({m_\ell })}}\sum\limits_{{j_\ell } = 0}^\infty  {\frac{{{K_\ell }^{{j_\ell }}{d_\ell }_{{j_\ell }}}}{{{j_\ell }!}}} \frac{2}{{\Gamma ({j_\ell } + 1)}}} \int_0^x {\frac{1}{t}G_{0,N}^{N,0}\left( {\left. {{t^2}\prod\limits_{\ell  = 1}^N {\left( {\frac{1}{{2{\sigma _\ell }^2}}} \right)} } \right|\begin{array}{*{20}{c}}
					- \\
					{1 + {j_1}, \cdots ,1 + {j_N}}
			\end{array}} \right){\rm d}t}.
	\end{equation}}
	Making the change of variable ${t^2} = a$ and using \cite[eq. (9.301)]{gradshteyn2007}, we can rewrite \eqref{CDFGUOCHENG1} as
	{\small\begin{equation}\label{CDFGUOCHENG}
	 {F_X}\left( x \right) = \prod\limits_{\ell  = 1}^N {\frac{{{m_\ell }^{{m_\ell }}}}{{\Gamma ({m_\ell })}}\sum\limits_{{j_\ell } = 0}^\infty  {\frac{{{K_\ell }^{{j_\ell }}{d_\ell }_{{j_\ell }}}}{{{j_\ell }!}}} \frac{1}{{\Gamma ({j_\ell } + 1)}}\frac{1}{{2\pi i}}} \int_{\cal L} {\prod\limits_{\ell  = 1}^N {\Gamma \left( {1 + {j_\ell } + t} \right)} {{\left( {\prod\limits_{\ell  = 1}^N {\left( {\frac{1}{{2{\sigma _\ell }^2}}} \right)} } \right)}^{ - t}}{I_{A_1}}{\rm d}t}
	\end{equation}}
	where
	{\small\begin{equation}\label{IARESULT}
	 {I_{A_1}}{{ = }}\int_0^{{x^2}} {{a^{ - 1 - t}}{\rm d}a}{{ = }}\frac{{\Gamma \left( { - t} \right)}}{{\Gamma \left( {1 - t} \right)}}{x^{ - {{2}}t}}
	\end{equation}}
where we have used \cite[eq. (8.331.3)]{gradshteyn2007}. Substituting \eqref{IARESULT} into \eqref{CDFGUOCHENG} and using \cite[eq. (9.301)]{gradshteyn2007}, we can obtain \eqref{CDFPRO}.
	
	\subsubsection{Proof of MGF}
	The generalized MGF of the product of FTR RVs can be obtained using
	{\small\begin{equation}\label{MGFCHUSHI1}	
	 {{\cal M}_X}\left( s \right) = \left[ {{e^{ - xs}}} \right] = \int_0^\infty  {{e^{ - xs}}{f_X}\left( x \right)} {\rm d}x.
	\end{equation}}
	Substituting \eqref{PDFFINAL1} into \eqref{MGFCHUSHI1} and then using \cite[eq. (9.301)]{gradshteyn2007}, we can rewrite \eqref{MGFCHUSHI1} as
		{\small\begin{equation}\label{MGFGUOCHENG1}
 {{\cal M}_X}\left( s \right)\! =\! \sum\limits_{{j_\ell } = 0}^\infty  {\prod\limits_{\ell  = 1}^N {\frac{{{K_\ell }^{{j_\ell }}{d_\ell }_{{j_\ell }}}}{{{j_\ell }!}}\frac{{{m_\ell }^{{m_\ell }}}}{{\Gamma ({m_\ell })}}\frac{2}{{\Gamma ({j_\ell } + 1)}}} } \frac{1}{{2\pi i}}\int_0^\infty  \!\!\!\!{{x^{ - 1}}} {e^{ - xs}}\!\!\int_{\cal L}\! {\prod\limits_{\ell  = 1}^N {\Gamma \left( {1 + {j_\ell } + t} \right)} {{\left( {{x^2}\prod\limits_{\ell  = 1}^N \!{\left( {\frac{1}{{2{\sigma _\ell }^2}}} \right)} } \!\right)}^{ - t}}} {\rm d}t{\rm d}x.
	\end{equation}}
	According to Fubini's theorem and \cite[eq. (3.381.4)]{gradshteyn2007}, we exchange the order of integrations in \eqref{MGFGUOCHENG1}, and derive
	{\small\begin{equation}\label{MGFA1}
	 {{\cal M}_X}\left( s \right){{ = }}\sum\limits_{{j_\ell } = 0}^\infty  {\prod\limits_{\ell  = 1}^N {\frac{{{K_\ell }^{{j_\ell }}{d_\ell }_{{j_\ell }}}}{{{j_\ell }!}}\frac{{{m_\ell }^{{m_\ell }}}}{{\Gamma ({m_\ell })}}\frac{2}{{\Gamma ({j_\ell } + 1)}}} } \frac{1}{{2\pi i}}\int_{\cal L} {\prod\limits_{\ell  = 1}^N {\Gamma \left( {1 + {j_\ell } + t} \right)} {{\left( {\prod\limits_{\ell  = 1}^N {\left( {\frac{1}{{2{\sigma _\ell }^2}}} \right)} } \right)}^{ - t}}} {I_{{A_2}}}{\rm d}t
	\end{equation}}
	where
		{\small \begin{equation}\label{IA2}
{I_{{A_2}}} = \int_0^\infty  {{x^{- 2t - 1}}} {e^{ - xs}}{\rm d}x= {s^{2t}}\Gamma \left( {q - 2t} \right).
	\end{equation}}
	With the help of definition of Fox's $H$-function \cite[eq. (1.2)]{mathai2009h}, we obtain \eqref{MGFPRO} to complete the proof.
	
	\section{Proof of Theorem \ref{sumrvftr}}\label{AB}
	\renewcommand{\theequation}{B-\arabic{equation}}
	\setcounter{equation}{0}
	
	\subsubsection{Proof of PDF}\label{SUBPDF}
	Let ${{{\cal M}_{X_\iota}}\left( s \right)} $ denote the MGF of the ${l_{{\rm th}}}$ product of two FTR RVs and we can derive the MGF of $Z$ as
		{\small\begin{equation}
 {{\cal M}_Z}\left( s \right) = \prod\limits_{\iota  = 1}^L {{{\cal M}_{X_\iota}}\left( s \right)} .
	\end{equation}			}
	Thus, the PDF of $X$ can be expressed as
	{\small\begin{equation}\label{PDF2GUOCHENG}	
	 {f_Z}(z) = {{\cal L}^{ - 1}}\left[ {{{\cal M}_X}\left( s \right);z} \right].
	\end{equation}}
	Substituting \eqref{MGFPRO} into \eqref{PDF2GUOCHENG}, after some algebraic manipulations, we have
	{\small\begin{equation}\label{PDFDETUIDAO}	
	 {f_Z}(z){{ = }}\sum\limits_{{j_{1.1}}, \cdots ,{j_{L,1}} = 0}^\infty   \cdots  \sum\limits_{{j_{1.N}}, \cdots ,{j_{L,N}} = 0}^\infty  {\prod\limits_{\iota  = 1}^L {\prod\limits_{\ell  = 1}^N {\frac{{{K_{\iota ,\ell }}^{{j_{\iota ,\ell }}}{d_{\iota ,\ell }}_{{j_{\iota ,\ell }}}}}{{{j_{\iota ,\ell }}!}}\frac{{{m_{\iota ,\ell }}^{{m_{\iota ,\ell }}}}}{{\Gamma ({m_{\iota ,\ell }})}}\frac{1}{{\Gamma ({j_{\iota ,\ell }} + 1)}}} {I_{{B_1}}}} }
	\end{equation}}
	where
	{\small\begin{align}	
	 {I_{{B_1}}}& = {{\cal L}^{ - 1}}\left[ {{{\left( {\frac{1}{{2\pi i}}} \right)}^\iota }\int_{{{\cal L}_\iota }} {\prod\limits_{\ell  = 1}^N {\Gamma \left( {1 + {j_{\iota ,\ell }} + \frac{1}{2}{\varsigma _\iota }} \right)} \Gamma \left( { - {\varsigma _\iota }} \right){{\left( {s\prod\limits_{\ell  = 1}^N {\left( {\sqrt 2 {\sigma _{\iota ,\ell }}} \right)} } \right)}^{{\varsigma _\iota }}}} {\rm d}{\varsigma _\iota };z} \right]\notag\\
		&{ = }\frac{1}{{2\pi i}}\int_{\cal L} {{{\left( {\frac{1}{{2\pi i}}} \right)}^\iota }\int_{{{\cal L}_\iota }} {\prod\limits_{\ell  = 1}^N {\Gamma \left( {1 + {j_{\iota ,\ell }} + \frac{1}{2}{\varsigma _\iota }} \right)} \Gamma \left( { - {\varsigma _\iota }} \right){{\left( {s\prod\limits_{\ell  = 1}^N {\left( {\sqrt 2 {\sigma _{\iota ,\ell }}} \right)} } \right)}^{{\varsigma _\iota }}}} {\rm d}{\varsigma _\iota }{e^{sz}}{\rm d}s} .
	\end{align}	}
	Note that the order of integration can be interchangeable, we can express $I_{B_1}$ as
	{\small\begin{equation}\label{IB1ZAIZHE}
	 {I_{{B_1}}}{{ = }}\frac{1}{{2\pi i}}{\left( {\frac{1}{{2\pi i}}} \right)^\iota }\int_{{{\cal L}_\iota }} {\prod\limits_{\ell  = 1}^N {\Gamma \left( {1 + {j_{\iota ,\ell }} + \frac{1}{2}{\varsigma _\iota }} \right)} \Gamma \left( { - {\varsigma _\iota }} \right){{\left( {\prod\limits_{\ell  = 1}^N {\left( {\sqrt 2 {\sigma _{\iota ,\ell }}} \right)} } \right)}^{{\varsigma _\iota }}}{I_{{B_2}}}} {\rm d}{\varsigma _\iota }
	\end{equation}}
	where
	{\small\begin{equation}
	 {I_{{B_2}}}{{ = }}\int_{\cal L} {{s^{\sum\limits_{\iota {{ = 1}}}^L {{\varsigma _\iota }} }}{e^{sz}}{\rm d}s} .
	\end{equation}}
	Let $ sz =  - b $ and we have
	{\small\begin{equation}
	 {I_{{B_2}}} =  - {\left( {\frac{1}{z}} \right)^{1 + \sum\limits_{\iota {{ = 1}}}^L {{\varsigma _\iota }} }}\int_{\cal L} {{{\left( { - b} \right)}^{\sum\limits_{\iota {{ = 1}}}^L {{\varsigma _\iota }} }}{e^{ - b}}} {\rm d}b.
	\end{equation}}
	With the aid of \cite[eq. (8.315.1)]{gradshteyn2007}, ${I_{{B_2}}}$ can be solved as
	{\small\begin{equation}\label{IB2ZAIZHELI}
	 {I_{{B_2}}} = {\left( {\frac{1}{z}} \right)^{1 + \sum\limits_{\iota {{ = 1}}}^L {{\varsigma _\iota }} }}\frac{{2\pi i}}{{\Gamma \left( { - \sum\limits_{\iota {{ = 1}}}^L {{\varsigma _\iota }} } \right)}}.
	\end{equation}}
	Combining \eqref{IB2ZAIZHELI}, \eqref{IB1ZAIZHE} and \eqref{PDFDETUIDAO} and using the definition of the multivariate Fox's $H$-function \cite[eq. (A-1)]{mathai2009h}, we obtain \eqref{PDFFTRSUMPRO} and complete the proof.
	\subsubsection{Proof of CDF}
	Following similar procedures as in Appendix \ref{SUBPDF}, we can derive the CDF of $Z$ by taking the inverse Laplace transform of $Mz (s) /s$.
	
	\section{Proof of Lemma \ref{ACL}}\label{AC}
	\renewcommand{\theequation}{C-\arabic{equation}}
	\setcounter{equation}{0}
	It is obvious that $ {F_{{\gamma _{{\rm{AF}}}}}}\left( z \right) = 0 $ for $ z < 0 $ and $ {F_{{\gamma _{{\rm{AF}}}}}}\left( z \right) = 1 $ for $ z > 1/d $. Therefore, hereafter it is assumed that $ 0 \le z \le 1/d $. Let us define $U=c_2/\gamma_1$, $V=c_1/\gamma_2$ and $W=1/\gamma_{AF}-d$. The PDF of W is obtained using the convolution theorem, namely
	{\small\begin{equation}
	 {F_W}(w) = \int_0^w {{F_U}} (w - x){f_V}\left( x \right){\rm d}x.
	\end{equation}}
	The CDFs of $U$ and $V$ can be expressed in terms of the CDFs of $z_1$ and $z_2$, respectively, as
	{\small\begin{equation}\label{jifenyangzi}
	 {F_U}\left( u \right){{ = }}1 - {F_{{\gamma _1}}}\left( {\frac{{{c_2}}}{u}} \right),\: {f_V}(v) = \frac{{{c_1}}}{{{v^2}}}{f_{{\gamma _2}}}\left( {\frac{{{c_1}}}{v}} \right).
	\end{equation}}
	Thus, eq. \eqref{jifenyangzi} can be expressed as
	{\small\begin{equation}
	 {F_W}(w) = \int_0^w {{f_{{\gamma _2}}}\left( {\frac{{{c_1}}}{x}} \right)\frac{{{c_1}}}{{{x^2}}}} dx - \int_0^w {{F_{{\gamma _1}}}\left( {\frac{{{c_2}}}{{w - x}}} \right)} {f_{{\gamma _2}}}\left( {\frac{{{c_1}}}{x}} \right)\frac{{{c_1}}}{{{x^2}}}{\rm d}x.
	\end{equation}}
	By performing the change of variables $x=wt$, ${F_W}(w)$ can be further expressed as
	{\small\begin{equation}\label{daoshuer}
	 {F_W}(w) = \int_0^1 {{f_{{\gamma _2}}}\left( {\frac{{{c_1}}}{{wt}}} \right)\frac{{{c_1}}}{{w{t^2}}}} {\rm d}t - \int_0^1 {{F_{{\gamma _1}}}\left( {\frac{{{c_2}}}{{w\left( {1 - t} \right)}}} \right)} {f_{{\gamma _2}}}\left( {\frac{{{c_1}}}{{wt}}} \right)\frac{{{c_1}}}{{w{t^2}}}{\rm d}t.
	\end{equation}}
	Noticing that $ {Z} = \frac{1}{W+d} $ and employing a transformation of RVs, the CDF of $\gamma_{AF}$ can be derived in terms of the CDF of $W$ as
	{\small\begin{equation}\label{afhew}
	 {F_{{Z}}}\left( z \right) = 1 - {F_W}\left(\frac{1}{z} - d\right).
	\end{equation}}
	With the help of \eqref{afhew} and \eqref{daoshuer}, we obtain \eqref{zhuizhongCDF} and complete the proof.
	\section{Proof of Theorem \ref{them4}}\label{AT4}
	\renewcommand{\theequation}{D-\arabic{equation}}
	\setcounter{equation}{0}
	\subsubsection{Proof of PDF}
	Substituting \eqref{PDFFTR} to \eqref{jifenafpdf}, after some straightforward manipulations, we obtain
	{\small\begin{equation}
	 {f_{{\gamma _{AF}}}}\left( z \right)=\frac{{{c_2}{c_1}z}}{{{{(1 - zd)}^3}}}\sum\limits_{{j_{{1}}} = 0}^\infty  {\sum\limits_{{j_{{2}}} = 0}^\infty  {\prod\limits_{\ell  = 1}^2 {\left( {\frac{{{m_\ell }^{{m_\ell }}}}{{\Gamma ({m_\ell })}}\frac{{{K_\ell }^{{j_\ell }}{d_\ell }_{{j_\ell }}}}{{{j_\ell }!}}\frac{{{A_\ell }^{{j_\ell }}}}{{2{\sigma _\ell }^2\Gamma ({j_\ell } + 1)}}} \right)} } } {I_{{C_1}}}
	\end{equation}}
	where
	{\small\begin{equation}
	 {I_{{C_1}}} = \int_0^1 {\frac{1}{{{t^{2 + {j_1}}}}}} \frac{{{1}}}{{{{\left( {1 - t} \right)}^{2 + {j_2}}}}}\exp \left( { - \frac{{{1}}}{{2{\sigma _{{1}}}^2}}\frac{{{c_2}z}}{{(1 - zd)t}}} \right)\exp \left( { - \frac{{{1}}}{{2{\sigma _{{2}}}^2}}\frac{{{c_1}z}}{{(1 - zd)(1 - t)}}} \right){\rm{d}}t.
	\end{equation}}
	By using \cite[eq. (01.03.07.0001.01)]{web} and exchanging the order of integrations, we can express ${I_{{C_1}}}$ as
	{\small\begin{equation}\label{ICzah}
	 {I_{{C_1}}} = \int_{{{\cal L}_1}} {\int_{{{\cal L}_2}} {\Gamma \left( {{s_1}} \right)\Gamma \left( {{s_2}} \right)} } {\left( {\frac{{{1}}}{{2{\sigma _{{1}}}^2}}\frac{{{c_2}z}}{{(1 - zd)}}} \right)^{ - {s_1}}}{\left( {\frac{{{1}}}{{2{\sigma _{{2}}}^2}}\frac{{{c_1}z}}{{(1 - zd)}}} \right)^{ - {s_2}}}{I_{{C_2}}}{\rm d}{s_2}{\rm d}{s_1}
	\end{equation}}
	where
	{\small\begin{equation}`
	 {I_{{C_2}}} = \int_0^1 {\frac{1}{{{t^{2 + {j_1} - {s_1}}}}}} \frac{{{1}}}{{{{\left( {1 - t} \right)}^{2 + {j_2} - {s_2}}}}}{\rm{d}}t.
	\end{equation}}
	With the help of \cite[eq. (3.251.1)]{gradshteyn2007} and \cite[eq. (8.384.1)]{gradshteyn2007}, ${I_{{C_2}}}$ can be solved as
	{\small\begin{equation}\label{ic2zaizhe}
	 {I_{{C_2}}} = \frac{{\Gamma \left( {1 - 2 - {j_1} + {s_1}} \right)\Gamma \left( {1 - 2 - {j_{{2}}} + {s_{{2}}}} \right)}}{{\Gamma \left( { - 2 - {j_{{2}}} + {s_{{2}}} - {j_1} + {s_1}} \right)}}.
	\end{equation}}
	Combining \eqref{ic2zaizhe}, \eqref{ICzah} and using the definition of multivariate Fox's $H$-function \cite[eq. (A-1)]{mathai2009h}, we obtain \eqref{PDFAF} to complete the proof.
	\subsubsection{Proof of CDF}	
	Substituting \eqref{PDFFTR} and \eqref{CDFFTR} to \eqref{zhuizhongCDF}, we have
	{\small\begin{equation}
	 {F_{{\gamma _{{{AF}}}}}}\left( z \right) = 1 - F_{{1}} + F_{{2}}
	\end{equation}}
	where
	{\small\begin{equation}
	 {F_{{1}}} = \frac{{z{c_1}}}{{\left( {1 - zd} \right)}}\frac{{{m_{{2}}}^{{m_{{2}}}}}}{{\Gamma ({m_{{2}}})}}\sum\limits_{{j_{{2}}} = 0}^\infty  {\frac{{{K_{{2}}}^{{j_{{2}}}}{d_{{2}}}_{{j_{{2}}}}}}{{{j_{{2}}}!}}} \frac{{{{\left( {\frac{{z{c_1}}}{{\left( {1 - zd} \right)}}} \right)}^{{j_{{2}}}}}}}{{\Gamma ({j_{{2}}} + 1){{\left( {2{\sigma _{{2}}}^2} \right)}^{{j_2} + 1}}}}\int_0^1 {\frac{{{1}}}{{{t^{{j_2}}}}}\exp \left( { - \frac{1}{{2{\sigma _{{2}}}^2}}\frac{{z{c_1}}}{{\left( {1 - zd} \right)t}}} \right)} \frac{{{\rm d}t}}{{{t^2}}},
	\end{equation}}
	{\small\begin{equation}\label{F2}
	 {F_{{2}}} = \frac{{z{c_1}}}{{\left( {1 - zd} \right)}}\frac{{{m_{{1}}}^{{m_{{1}}}}}}{{\Gamma ({m_{{1}}})}}\frac{{{m_{{2}}}^{{m_{{2}}}}}}{{\Gamma ({m_{{2}}})}}\sum\limits_{{j_{{1}}} = 0}^\infty  {\sum\limits_{{j_{{2}}} = 0}^\infty  {\frac{{{K_{{1}}}^{{j_{{1}}}}{d_1}_{{j_{{1}}}}}}{{{j_{{1}}}!}}\frac{{{K_{{2}}}^{{j_{{2}}}}{d_{{2}}}_{{j_{{2}}}}}}{{{j_{{2}}}!}}} } \frac{1}{{\Gamma ({j_{{1}}} + 1)}}\frac{{{{\left( {\frac{{z{c_1}}}{{\left( {1 - zd} \right)}}} \right)}^{{j_{{2}}}}}}}{{\Gamma ({j_{{2}}} + 1){{\left( {2{\sigma _{{2}}}^2} \right)}^{{j_2} + 1}}}}{I_{{C_3}}}
	\end{equation}}
	where
	{\small\begin{equation}
	 {I_{{C_3}}} = \int_0^1 {\frac{1}{{{t^{{j_2} + 2}}}}} \exp \left( { - \frac{1}{{2{\sigma _{{2}}}^2}}\frac{{z{c_1}}}{{\left( {1 - zd} \right)t}}} \right)\gamma \left( {{j_{{1}}} + 1,\frac{{{1}}}{{2{\sigma _{{1}}}^2}}\frac{{z{c_2}}}{{\left( {1 - zd} \right)\left( {1 - t} \right)}}} \right){\rm d}t.
	\end{equation}}
	With the help of \cite[eq. (3.381.8)]{gradshteyn2007}, $F_1$ can be solved as
	{\small\begin{equation}
	 {F_{{1}}} = \frac{{{m_{{2}}}^{{m_{{2}}}}}}{{\Gamma ({m_{{2}}})}}\sum\limits_{{j_{{2}}} = 0}^\infty  {\frac{{{K_{{2}}}^{{j_{{2}}}}{d_{{2}}}_{{j_{{2}}}}}}{{{j_{{2}}}!}}} \frac{1}{{\Gamma ({j_{{2}}} + 1)}}\Gamma \left( {1 + {j_2},\frac{1}{{2{\sigma _{{2}}}^2}}\frac{{z{c_1}}}{{\left( {1 - zd} \right)}}} \right).
	\end{equation}}
	Using \cite[eq. (01.03.07.0001.01)]{web}, \cite[eq. (06.06.07.0002.01)]{web}, \cite[eq. (3.191.1)]{gradshteyn2007} and \cite[eq. (8.384.1)]{gradshteyn2007} we can express $I_{C_3}$ as
	{\small\begin{equation}\label{icsan}
	 {I_{{C_3}}} = {\left(\! {\frac{{{1}}}{{{{2}}\pi i}}}\! \right)^{{2}}}\!\!\int_{{{\cal L}_{{1}}}} {\int_{{{\cal L}_{{2}}}} {\Gamma \left( {{s_{{2}}}} \right)\frac{{\Gamma \left( {{s_1} + {j_{{1}}} + 1} \right)\Gamma \left( { - {s_1}} \right)}}{{\Gamma \left(\! {{{1}} - {s_1}} \!\right)}}} {{\left( {\frac{1}{{2{\sigma _{{2}}}^2}}\frac{{z{c_1}}}{{\left( {1 - zd} \right)}}} \right)}^{ - {s_2}}}} {\left(\! {\frac{{{1}}}{{2{\sigma _{{1}}}^2}}\frac{{z{c_2}}}{{\left( {1 - zd} \right)}}}\! \right)^{ - {s_1}}}{I_{{C_4}}}{\rm d}{s_2}{\rm d}{s_1}
	\end{equation}}
	where
	{\small\begin{equation}\label{ic4}
	 {I_{{C_4}}} = \int_0^1 {\frac{1}{{{t^{{j_2} + 2 - {s_2}}}{{\left( {1 - t} \right)}^{ - {s_1}}}}}} {\rm d}t= \frac{{\Gamma \left( {1 + {s_1}} \right)\Gamma \left( {{s_2} - {j_2} - 1} \right)}}{{\Gamma \left( {{s_1} + {s_2} - {j_2}} \right)}}.
	\end{equation}}
	Combining \eqref{ic4}, \eqref{icsan}, \eqref{F2} and using the definition of multivariate Fox's $H$-function \cite[eq. (A-1)]{mathai2009h}, we obtain \eqref{CDFAF} and complete the proof.
	\section{Proof of Lemma \ref{consd}}\label{AppendixEE}
	\renewcommand{\theequation}{E-\arabic{equation}}
	\setcounter{equation}{0}
{\color{black}	When $k\ge1$, we can express ${\alpha _{kL + \ell }}$ as
	{\small\begin{equation}\label{afdagb}
	 {\alpha _{kL + \ell }} = \frac{1}{{L - 1}}\left( {{S_{kL + \ell  - 1}} - {S_{kL + \ell  - L}}} \right)
	\end{equation}}
	where ${S_\ell } = \sum\limits_{i = 1}^\ell  {{\alpha _\ell }}$. Thus, we obtain
	{\small\begin{equation}\label{Afevd}
	 {\alpha _{kL + \ell  + 1}} - {\alpha _{kL + \ell }} = \frac{1}{{L - 1}}\left( {{S_{kL + \ell }} - {S_{kL + \ell  + 1 - L}} - {S_{kL + \ell  - 1}} + {S_{kL + \ell  - L}}} \right) = \frac{1}{{L - 1}}\left( {{\alpha _{kL + \ell }} - {\alpha _{kL + \ell  - L}}} \right).
	\end{equation}}
	After some algebraic manipulations, eq. \eqref{Afevd} can be rewritten as
	{\small\begin{equation}\label{dafa}
	 \frac{{{\alpha _{kL + \ell  + 1}}}}{{{\alpha _{kL + \ell }}}} = \frac{L}{{L - 1}} - \frac{1}{{L - 1}}\frac{{{\alpha _{kL + \ell  - L}}}}{{{\alpha _{kL + \ell }}}} = \frac{L}{{L - 1}} - \frac{1}{{L - 1}}\frac{{{\alpha _{kL + \ell  - L}}}}{{{\alpha _{kL + \ell  - L + 1}}}} \cdots \frac{{{\alpha _{kL + \ell  - 1}}}}{{{\alpha _{kL + \ell }}}}.
	\end{equation}}
	We define that $ \mathop {\lim }\limits_{k \to \infty } \frac{{{\alpha _{kL + \ell  + 1}}}}{{{\alpha _{kL + \ell }}}}=  C $. Substituting $C$ into \eqref{dafa}, we can obtain
	{\small\begin{equation}\label{faljf}
	 C = \frac{L}{{L - 1}} - \frac{1}{{L - 1}}\frac{{{1}}}{{{C^L}}}.
	\end{equation}}
	Obviously, for any amount of reflecting elements on the RIS, $L$, we can solve \eqref{faljf} and obtain $C=1$. Therefore, we confirm that $ \mathop {\lim }\limits_{k \to \infty } {\alpha _{kL + \ell }} = \phi $. To obtain $\phi$, with the help of \eqref{afdagb}, we derive
	{\small\begin{equation}\label{ffafdagb}
	 {\alpha _{kL + 1}} +  \cdots  + {\alpha _{kL + L}} = \frac{1}{{L - 1}}\left( {{\alpha _{\left( {k - 1} \right)L + 2}} +  \cdots  + {\alpha _{\left( {k - 1} \right)L + L}} + {\alpha _{kL + 1}} + {\alpha _{\left( {k - 1} \right)L + {\rm{3}}}} +  \cdots } \right).
	\end{equation}}
	After some algebraic manipulations, eq. \eqref{ffafdagb} can be expressed as
	{\small\begin{align}\label{afeag}
	 {\alpha _{kL + 2}} + 2{\alpha _{kL + 3}} +  \cdots  + \left( {L - 1} \right){\alpha _{kL + L}} &= {\alpha _{\left( {k - 1} \right)L + 2}} + 2{\alpha _{\left( {k - 1} \right)L + {{3}}}} +  \cdots  + \left( {L - 1} \right){\alpha _{\left( {k - 1} \right)L + L}}\notag\\
		& = {\alpha _2} + 2{\alpha _3} + 3{\alpha _4} \cdots  + \left( {L - 1} \right){\alpha _L}.
	\end{align}}
	Substituting $ \mathop {\lim }\limits_{k \to \infty } {\alpha _{kL + \ell }} = \phi $ into \eqref{afeag}, we derive
	{\small\begin{equation}\label{aefav}
	 \left( {1 + 2 +  \cdots  + \left( {L - 1} \right)} \right)\phi  = \frac{{L\left( {L - 1} \right)}}{2}\phi  = {\alpha _2} + 2{\alpha _3} + 3{\alpha _4} \cdots  + \left( {L - 1} \right){\alpha _L}.
	\end{equation}}
	Thus, with the help of \eqref{aefav}, \eqref{afagbv} can be obtained which completes the proof.}
	
	\section{Proof of Theorem \ref{sfagea}}\label{AppendixE}
	\renewcommand{\theequation}{F-\arabic{equation}}
	\setcounter{equation}{0}
	Substituting \eqref{CDFFTRSUMPRO} in \eqref{BERDINGYI} and changing the order of integration, we can obtain
	{\small \begin{align}	\label{Peris}
	{P_e} = &\sum\limits_{{j_{1.1}}, \cdots ,{j_{L,1}} = 0}^\infty   \cdots  \sum\limits_{{j_{1,N}}, \cdots ,{j_{L,N}} = 0}^\infty  {\prod\limits_{\iota  = 1}^L {\prod\limits_{\ell  = 1}^N {\frac{{{K_{\iota ,\ell }}^{{j_{\iota ,\ell }}}{d_{\iota ,\ell }}_{{j_{\iota ,\ell }}}}}{{{j_{\iota ,\ell }}!}}\frac{{{m_{\iota ,\ell }}^{{m_{\iota ,\ell }}}}}{{\Gamma ({m_{\iota ,\ell }})}}\frac{1}{{\Gamma ({j_{\iota ,\ell }} + 1)}}} } }
	\notag\\
	&\times \prod\limits_{\iota  = 1}^L {\frac{1}{{2\pi i}}\int_{{{\cal L}_\iota }} {\frac{{\prod\limits_{\ell  = 1}^N {\Gamma \left( {1 + {j_{\iota ,\ell }} + \frac{1}{2}{\varsigma _\iota }} \right)} \Gamma \left( { - {\varsigma _\iota }} \right)}}{{\Gamma \left( {1 - \sum\limits_{\iota {{ = 1}}}^L {{\varsigma _\iota }} } \right)}}{{\left( {{{\left( {\frac{P}{{{o^2}}}} \right)}^{\frac{1}{2}}}}{\prod\limits_{\ell  = 1}^N {\left( {\sqrt 2 {\sigma _{\iota ,\ell }}} \right)} } \right)}^{{\varsigma _\iota }}}} {\rm d}{\varsigma _\iota }} \frac{{{q^p}}}{{2\Gamma (p)}}{I_{{D_1}}}
	\end{align}}
	With the help of \cite[eq. (3.351.3)]{gradshteyn2007} and \cite[eq. (8.331.1)]{gradshteyn2007}, ${I_{{D_1}}}$ can be expressed as
	{\small \begin{equation}\label{id1zuihou}
			{I_{{D_1}}}= \int_0^\infty  {{z^{ - 0.5\sum\limits_{\iota {{ = 1}}}^N {{\varsigma _\iota }}  + p - 1}}} {e^{ - qz}}{\rm d}z= {q^{ - p + 0.5\sum\limits_{\iota {{ = 1}}}^N {{\varsigma _\iota }} }}\Gamma \left( {p -0.5 \sum\limits_{\iota {{ = 1}}}^N {{\varsigma _\iota }} } \right).
	\end{equation}}
	Letting $N=2$ and substituting \eqref{id1zuihou} in \eqref{Peris}, we obtain {\small \begin{align}\label{sds}
	{P_e} = &\frac{{{1}}}{{2\Gamma (p)}}\sum\limits_{{j_{1.1}}, \cdots ,{j_{L,1}} = 0}^\infty   \cdots  \sum\limits_{{j_{1,N}}, \cdots ,{j_{L,N}} = 0}^\infty  {\prod\limits_{\iota  = 1}^L {\prod\limits_{\ell  = 1}^N {\frac{{{K_{\iota ,\ell }}^{{j_{\iota ,\ell }}}{d_{\iota ,\ell }}_{{j_{\iota ,\ell }}}}}{{{j_{\iota ,\ell }}!}}\frac{{{m_{\iota ,\ell }}^{{m_{\iota ,\ell }}}}}{{\Gamma ({m_{\iota ,\ell }})}}\frac{1}{{\Gamma ({j_{\iota ,\ell }} + 1)}}} } }
	\notag\\&\times
	\prod\limits_{\iota  = 1}^L {\frac{1}{{2\pi i}}\int_{{{\cal L}_\iota }} {\prod\limits_{\ell  = 1}^2 {\Gamma \left( {1 + {j_{\iota ,\ell }} + \frac{1}{2}{\varsigma _\iota }} \right)} \Gamma \left( { - {\varsigma _\iota }} \right)\frac{{\Gamma \left( {p - 0.5\sum\limits_{\iota {{ = 1}}}^N {{\varsigma _\iota }} } \right)}}{{\Gamma \left( {1 - \sum\limits_{\iota {{ = 1}}}^L {{\varsigma _\iota }} } \right)}}{{\left( {{{\left( {\frac{qP}{{{o^2}}}} \right)}^{\frac{1}{2}}}\prod\limits_{\ell  = 1}^2 {\left( {\sqrt 2 {\sigma _{\iota ,\ell }}} \right)} } \right)}^{{\varsigma _\iota }}}} {\rm d}{\varsigma _\iota }}.
	\end{align}}
	With the help of \cite[eq. (A-1)]{mathai2009h}, eq. \eqref{sds} can be expressed as \eqref{BERRISZUIZHONG} which completes the proof.
	\section{Proof of Theorem \ref{sfdsff}}\label{AppendixF}
	\renewcommand{\theequation}{G-\arabic{equation}}
	\setcounter{equation}{0}
	Substituting \eqref{CDFAFOUTH} into \eqref{BERDINGYI}, we obtain
	{\small \begin{align}\label{PAF}
	P_e^{AF} =& \frac{{{q^p}}}{{2\Gamma (p)}}{I_{E_1}} - \frac{{{q^p}}}{{2\Gamma (p)}}\frac{{{m_{2}}^{{m_{2}}}}}{{\Gamma ({m_{2}})}}\sum\limits_{{j_{2}} = 0}^\infty  {\frac{{{K_{2}}^{{j_{{2}}}}{d_{2}}_{{j_{2}}}}}{{{j_{2}}!}}} \frac{1}{{\Gamma ({j_{2}} + 1)}}{I_{E_2}}
	\notag\\&
	+ \frac{{{q^p}}}{{2\Gamma (p)}}\sum\limits_{{j_{1}} = 0}^\infty  {\sum\limits_{{j_{2}} = 0}^\infty  {\prod\limits_{\ell  = 1}^2 {\left( {\frac{{{m_\ell }^{{m_\ell }}}}{{\Gamma ({m_\ell })}}\frac{{{K_\ell }^{{j_\ell }}{d_\ell }_{{j_\ell }}}}{{{j_\ell }!}}\frac{{1}}{{\Gamma ({j_\ell } + 1)}}} \right)} } } {\left( {\frac{1}{{2{\sigma _{2}}^2}}} \right)^{{j_{2}} + 1}}{\left( {\frac{{1}}{{{2}\pi i}}} \right)^{2}}\int_{{{\cal L}_{1}}} {\int_{{{\cal L}_{2}}}}{I_{E_3}}
	\notag\\
	&\times {\frac{{\Gamma \left( {{s_1} + {j_{1}} + 1} \right)\Gamma \left( {1 + {s_1}} \right)\Gamma \left( { - {s_1}} \right)}}{{\Gamma \left( {{1} - {s_1}} \right)}}} \frac{{\Gamma \left( {{s_{2}}} \right)\Gamma \left( {{s_2} - {j_2} - 1} \right)}}{{\Gamma \left( {{s_1} + {s_2} - {j_2}} \right)}}{{\left( {\frac{o^2}{{2P_2{\sigma _2}^2}}} \right)}^{ - {s_2}}} {\left( {\frac{o^2}{{2P_1{\sigma _1}^2}}} \right)^{ - {s_1}}}{\rm d}{s_2}{\rm d}{s_1}
	\end{align}}
	where
	{\small \begin{equation}
			{I_{E_1}} = \int_0^\infty  {{z^{p - 1}}} {e^{ - qz}}{\rm d}z,
	\end{equation}}
	{\small \begin{equation}
			{I_{E_2}} = \int_0^\infty  {{z^{p - 1}}} {e^{ - qz}}\Gamma \left( {1 + {j_2},\frac{o^2z}{{2P_2{\sigma _2}^2}}} \right){\rm d}z,
	\end{equation}}
	and
	{\small \begin{equation}
			{I_{E_3}} = \int_0^\infty  {{z^{p - {s_1} - {s_2} + {j_2}}}} {e^{ - qz}}{\rm d}z.
	\end{equation}}
	Following similar procedures as \eqref{id1zuihou}, using \cite[eq. (3.351.3)]{gradshteyn2007} and \cite[eq. (8.331.1)]{gradshteyn2007}, we can solve $I_{E_1}$ and $I_{E_3}$ respectively as
	{\small \begin{equation}\label{P1zheli}
			{I_{E_1}} = {q^{ - p}}\Gamma \left( p \right),
			{I_{E_3}} = {q^{ - 1 - {j_2} - p + {s_1} + {s_2}}}\Gamma \left( {1 + {j_2} + p - {s_1} - {s_2}} \right).
	\end{equation}}
	With the help of \cite[eq. (6.455.1)]{gradshteyn2007}, ${I_{E_2}}$ can be solved as
	{\small \begin{equation}\label{P2ZHELI}
			{I_{E_2}}  = \frac{{\Gamma \left( {p + 1 + {j_2}} \right){{\left( {\frac{{2{P_2}{\sigma _2}^2}}{{{o^2}}}} \right)}^p}}}{{p{{\left( {{{1}} + \frac{{2{P_2}{\sigma _2}^2}}{{{o^2}}}q} \right)}^{p + 1 + {j_2}}}}}{}_2{F_1}\left( {1,p + 1 + {j_2};p + 1;\frac{q}{{\frac{1}{{2{\sigma _2}^2}} + q}}} \right).
	\end{equation}}
	Substituting \eqref{P1zheli} and \eqref{P2ZHELI} into \eqref{PAF}, we derive \eqref{BERAF} to complete the proof.
	
	\end{appendices}	
	\bibliographystyle{IEEEtran}
	\bibliography{IEEEabrv,Ref}
	\end{document}